\documentclass[twocolumn,pre, superscriptaddress, showpacs]{revtex4}
\usepackage{hyperref,xcolor}
\usepackage{amsmath,amssymb}
\usepackage{wrapfig}
\usepackage{subfig}
\usepackage{graphicx}
\usepackage{xcolor}

\newcommand{\imag}{\ensuremath{\mathrm{i}}}

\newcommand\Dfrtl[1]{\ensuremath{\,\mathrm{d}#1\,}}
\renewcommand\epsilon{\varepsilon}

\newcommand\euler[1]{\ensuremath{\mathrm{e}^{#1}}}

\begin{document}


\title{
Simulation of inhomogeneous distributions of ultracold atoms in an optical lattice via a massively 
parallel implementation of nonequilibrium strong-coupling perturbation theory
}
\date{\today}
\author{Andreas Dirks}
\email{andreas@physics.georgetown.edu}
\affiliation{Department of Physics, Georgetown University, Washington, DC 20057, USA}
\author{Karlis Mikelsons}
\affiliation{Department of Physics, Georgetown University, Washington, DC 20057, USA}
\author{H.R. Krishnamurthy}
\affiliation{Centre for Condensed Matter Theory, Department of Physics, \\
Indian Institute of Science, Bangalore 560012, India, and \\
Jawaharlal Nehru Centre for Advanced Scientific Research, Bangalore 560064, India
}
\author{James K. Freericks}
\affiliation{Department of Physics, Georgetown University, Washington, DC 20057, USA}

\begin{abstract}
We present a nonequilibrium strong-coupling approach to inhomogeneous systems of ultracold atoms in optical lattices. We 
demonstrate its application to the Mott-insulating phase of a two-dimensional Fermi-Hubbard model in the presence of a 
trap potential.
Since the theory is formulated self-consistently,
the numerical implementation relies on a massively parallel evaluation of the self-energy and the Green's function at each lattice site, 
employing thousands of CPUs.
While the computation of the self-energy is straightforward to parallelize, the evaluation of the Green's function requires the
inversion of a large sparse $10^d\times 10^d$ matrix, with $d > 6$. 
As a crucial ingredient, our solution heavily relies on the smallness of the hopping as compared to the interaction strength and yields
a widely scalable realization of a rapidly converging iterative algorithm which evaluates all elements of the Green's function.
Results are validated by comparing with the homogeneous case via the local-density approximation.
These calculations also show that the local-density approximation is valid in non-equilibrium setups without
mass transport.
\end{abstract}

\maketitle

\section{Introduction}
The field of ultracold atoms in optical lattices has been a promising new opportunity for studying many-body effects which are 
important for condensed-matter physics in controlled environments \cite{blochreview, Esslinger2010}. In particular, fermionic atoms such as $^{40}$K may provide a 
direct path towards a "quantum simulation" of the Hubbard model which itself has a paradigmatic role in condensed matter physics
and is a key in understanding phenomena such as high-temperature superconductivity and strongly correlated magnetism. In these experiments, 
some novel possibilities to study physical correlations between constituents of the model are being explored.

In many cases, such experiments\cite{Stoeferle2004, Joerdens2008, Greif2011} drive the systems substantially beyond thermal equilibrium, so that they are inaccessible to
methods of conventional equilibrium or linear-response theory. Usually, one also encounters spatially inhomogeneous situations,
since the atoms in the optical lattice are being held in a trap potential which co-exists with the lattice potential.
In one-dimensional systems, many opportunities to provide computational benchmarks for such experiments exist, such as via the
density-matrix renormalization group \cite{Clark2004, Kollath2005, Kollath2006, Winkler2006, Daley2008}.
However, it is a challenging problem to describe two- and three-dimensional systems out of thermal equilibrium, especially when
they are also inhomogeneous.

In this paper we present a nonequilibrium strong-coupling approach to inhomogeneous systems of ultracold atoms in optical lattices.
The paper is structured as follows.
Section \ref{sec:model} discusses the Hubbard model for an optical lattice in a trap.
In Section \ref{sec:Formalism}, we outline the strong-coupling approach which enables us to simulate inhomogeneous higher-dimensional 
Hubbard systems out of equilibrium. In Section \ref{sec:algo}, we develop the massively parallel algorithm which is used to solve the resulting equations on a supercomputer. Section \ref{sec:results} presents results of the algorithm for the example of a modulated 
lattice depth and validates them by comparing to the previously introduced strong-coupling method for homogeneous systems\cite{Mikelsons2012} 
within the local-density approximation (LDA). Conclusions are given in Section \ref{sec:conclusion}.

\section{Model}
\label{sec:model}
We consider a Fermi Hubbard model in the presence of a trap potential,
i.e. 
\begin{equation}
H(t) = \mathcal{H}_0(t) - \sum_{i,j,\sigma}  J_{ij}(t) c_{i,\sigma}^\dagger c_{j,\sigma},
\label{eq:model}
\end{equation}
with
\begin{equation}
\begin{split}
\mathcal{H}_0(t) =& \sum_{i} \mathcal{H}^{(i)}_{0}(t) \\
=& \sum_{i, \sigma} \epsilon_i(t)n_{i,\sigma} + \sum_i U_i(t) n_{i\uparrow} n_{i\downarrow},
\end{split}
\end{equation}
where the on-site single-particle energy levels
\begin{equation}
\epsilon_i(t) = V_\text{trap}(\vec r_i; t) - \frac{U_i(0)}2 - \mu
\label{eq:onsitelevels}
\end{equation}
are determined by the trap potential $V_\text{trap}(\vec r_i; t)$ and a global chemical potential $\mu$ which characterizes 
the initial equilibrium state at time $t=0$. The initial state is assumed to have a temperature $k_B T=\beta^{-1}$. 
The time-dependent interaction $U_i(t)$ and the time-dependent hopping $J_{ij}(t)$ are chosen to be results of a 
tight-binding calculation for maximally localized Wannier functions computed from the translationally invariant case.
In the future, we plan to include corrections to the tight-binding parameters which result from generalized Wannier functions
for the inhomogeneous problem.

We assume the system to be in thermal equilibrium at time $t=0$ and to be driven out of equilibrium subsequently by the time-dependence
of the model parameters.

\section{Formalism}
\label{sec:Formalism}
We employ a second-order self-consistent expression for the self-energy \cite{Mikelsons2012} around the atomic limit which
is described by $\mathcal{H}_0$. The self-consistency takes advantage of a resummation of diagrams which yields a better approximation.
\begin{figure}
\includegraphics[width=0.9\linewidth]{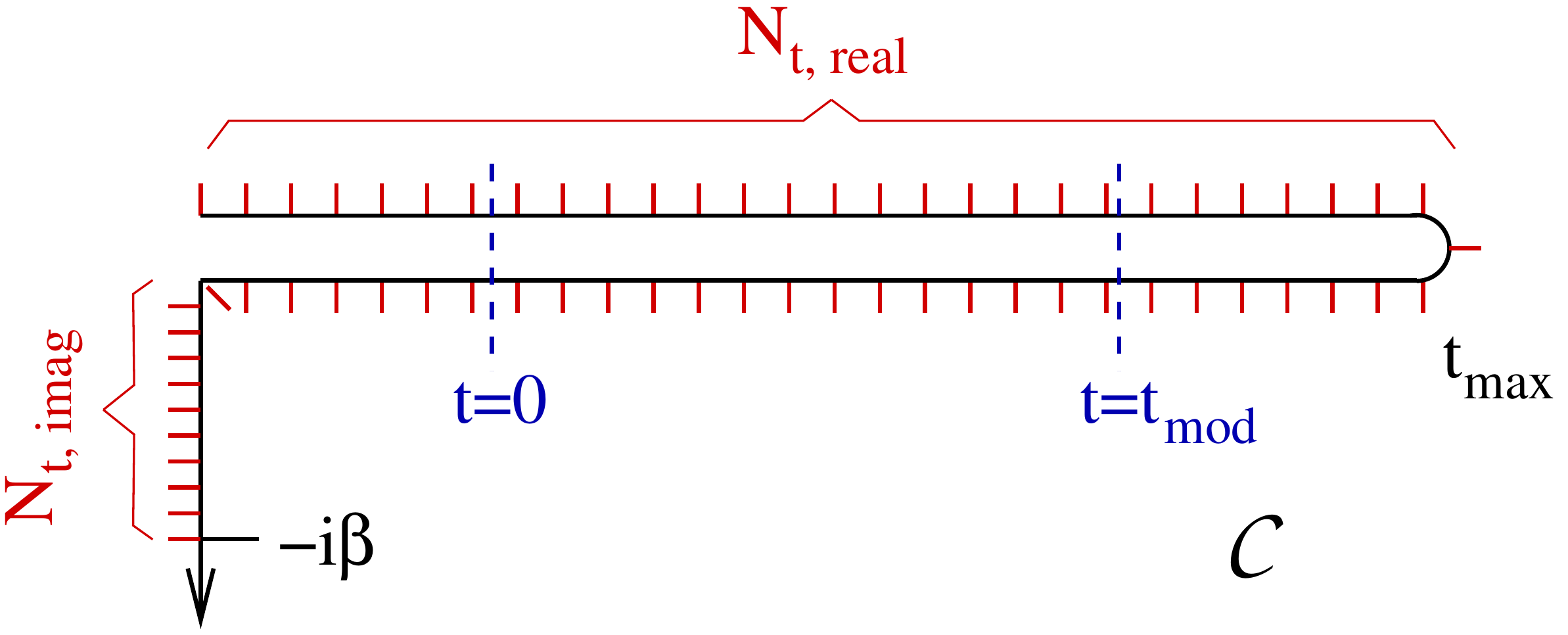}
\caption{(color online) Kadanoff-Baym-Keldysh contour $\mathcal{C}$ for the simulation time. Between the dashed blue lines, the system is driven out of
equilibrium by a time-dependent Hamiltonian. In the numerical part of the paper, we refer to the two points in time as $t=0$ and $t=t_\text{mod}$
for simplicity.
The real times between the Matsubara branch and the point at which the system is driven out of equilibrium can be used to check for numerical
convergence, since expectation values of observables have to be constant here. The time discretization is shown in red --- in the implementation
itself, the point $t=0$ is identical to the point to which the Matsubara branch of the contour is attached.
The total number of time slices is $N_t=2N_{t,\text{real}} + N_{t,\text{imag}}$.
}
\label{fig:Keldysh}
\end{figure}
It can be extended to an inhomogeneous system
in the following way:
\begin{equation}
\begin{split}
\Sigma_{lm,\sigma}(t,t') =& -\delta_{lm}\sum_{j_1,\sigma_1}
\int \Dfrtl{\tilde t}\int \Dfrtl{t_3} \int \Dfrtl{t_4} \int \Dfrtl{\tilde t'}
 \\
&\qquad\quad \times \mathcal{G}^{-1}_{l\sigma}(t,\tilde t) \,\tilde{\mathcal{G}}^{II}_{l;\sigma\sigma_1}(\tilde t, t_4; t_3,\tilde t')\\
&\qquad\quad \times J_{lj_1}(t_3)\, G^{(loc)}_{j_1 \sigma_1}(t_3,t_4) \\
&\qquad\quad \times J_{j_1l}(t_4)\,\mathcal{G}^{-1}_{l\sigma} (\tilde t', t')\\
=:& \,\delta_{lm}  \Sigma_{l,\sigma}(t,t').
\end{split}
\label{eq:selfcons}
\end{equation}
Here,
\begin{equation}
\mathcal{G}_{l\sigma}(t,t') = -\imag  \langle T_\mathcal{C} c_{l\sigma}(t) c_{l\sigma}^\dagger(t') \rangle_{\mathcal{H}^{(l)}_0}
\end{equation}
is the contour-ordered on-site no-hopping Green's function at site $l$ with spin $\sigma$ (in the paramagnetic phase, the 
Green's function is independent of the spin $\sigma$). Times $t$ and $t'$ are located on the Kadanoff-Baym-Keldysh contour $\mathcal{C}$ depicted in Fig.~\ref{fig:Keldysh}. 
\begin{equation}
\begin{split}
{\mathcal{\tilde G}}_{m;\sigma\tilde\sigma}^{II}(t_1,t_2;t_1',t_2')=&\,{\mathcal{ G}}_{m;\sigma\tilde\sigma}^{II}(t_1,t_2;t_1',t_2') \\
+&\, \mathcal{G}_{m\sigma}(t_1,t_1') \mathcal{G}_{m\sigma}(t_2,t_2')\delta_{\sigma\tilde\sigma} \\
-&\, \mathcal{G}_{m\sigma}(t_1,t_2') \mathcal{G}_{m\tilde\sigma}(t_2,t_1')
\end{split}
\end{equation}
is the second-order cumulant for the on-site no-hopping two-particle propagator
\begin{equation}
\begin{split}
&\mathcal{G}_{m;\sigma\tilde\sigma}^{II}(t_1,t_2;t_1',t_2') = 
\\&\quad(-\imag)^2 \langle T_\mathcal{C}c_{m\sigma}(t_1) c_{m\tilde\sigma}(t_2)
c^\dagger_{m\tilde\sigma}(t_1')c^\dagger_{m\sigma}(t_2')   \rangle_{\mathcal{H}^{(m)}_0}
\end{split}
\end{equation}
at site $m$. In a typical numerical implementation, one cannot store this tensor for a reasonable grid of time slices. However, it is easy
to compute it on the fly in the particle-number basis
$\{|E_{0}^{(m)}(t)\rangle,|E^{(m)}_{1}(t)\rangle, |E^{(m)}_{2}(t)\rangle, |E^{(m)}_{3}(t)\rangle\} = \{|0\rangle_m, |\downarrow\rangle_m, |\uparrow\rangle_m, |\downarrow\uparrow\rangle_m\}$ by inserting the time evolutions
\begin{equation}
c_{m\sigma}^{(\dagger)}(t) = \euler{\imag \int_0^t \mathcal{H}^{(m)}_{0}(t')\Dfrtl{t'}} c_{m\sigma}^{(\dagger)} \euler{-\imag \int_0^t \mathcal{H}^{(m)}_{0}(t')\Dfrtl{t'}}
\label{eq:localtimeevolution}
\end{equation}
for each possible $T_\mathcal{C}$ time ordering of the creation and annihilation operators. 
Note that it is crucial to the applicability of the approach that the expression in Eq.~\eqref{eq:localtimeevolution} does not involve any 
time-ordered products, because $[\mathcal{H}^{(m)}_0(t), \mathcal{H}^{(m)}_0(t')]=0$ for any combination of $t,t'$.
The on-the-fly evaluation of the action of the
operators in the basis can be realized by a fast multiplication with and division by tabulated 
\begin{equation}
\zeta^{(m)}_{\nu}(t) = \euler{\imag \int_0^t E_{\nu}^{(m)}(t')\Dfrtl{t'}}
\label{eq:cumulanttabulation}
\end{equation}
values.
For example, when $t_1 > t_2 > t_1' > t_2'$,
\begin{equation}
\begin{split}
\mathcal{G}^{II}_{m;\uparrow\downarrow}(t_1,t_2;t_1',t_2') =& (-\imag)^2\frac{\euler{-\beta E_{0}^{(m)}(0)}}{Z^{(m)}}\\
&\quad \times
\frac{\zeta_0^{(m)}(t_1) \zeta_2^{(m)}(t_2)   }
{\zeta_2^{(m)}(t_1) \zeta_3^{(m)}(t_2) } \\
&\quad\times 
\frac{ \zeta_3^{(m)}(t_1') \zeta_2^{(m)}(t_2')  }
{ \zeta_2^{(m)}(t_1') \zeta_0^{(m)}(t_2')}
,
\end{split}
\end{equation}
with 
\begin{equation}
Z^{(m)} = \sum_{\nu = 0}^3 \euler{-\beta E^{(m)}_{\nu}(0)}.
\end{equation}
In the calculation, one requires the expressions for all possible time orderings.
The relation in Eq.~\eqref{eq:selfcons} is solved iteratively. At a given step of this procedure, the self-energy at site $i$ is
given by
\begin{equation}
\Sigma_{i,\sigma} (t,t') = \int_{\tilde t \tilde t'}\mathcal G_{i,\sigma}^{-1}(t,\tilde t)
                        \tilde \Sigma_{i,\sigma}(\tilde t, \tilde t')
                           \mathcal{G}^{-1}_{i,\sigma}(\tilde t', t'),
\label{eq:defnSigma}
\end{equation}
where 
\begin{equation}
\begin{split}
\tilde \Sigma_{i,\sigma} (\tilde t, \tilde t') = &
- \sum_{j_1,\sigma_1}\int_{t_3,t_4} 
\tilde{\mathcal G}^{II}_{i,\sigma\sigma_1}(\tilde t, t_4; t_3,\tilde t')J_{i,j_1}(t_3) \\
& \qquad\qquad\qquad
\times G^{(loc)}_{j_1,\sigma_1}(t_3,t_4) J_{j_1,i}(t_4).
\end{split}
\label{eq:defnSigmatilde}
\end{equation}
The local Green's function at site $j_1$, $G^\text{(loc)}_{j_1,\sigma_1}$ is
given by the $j_1$-th block-diagonal element of the lattice Green's function
\begin{equation}
\begin{split}
G^\text{(latt)}_\sigma(\vec r_i,t;\vec r_{i'},t') =&
  -\imag \langle
   T_\mathcal{C} c_{\vec r_i\sigma}(t) c^\dagger_{\vec r_{i'}\sigma}(t')
  \rangle \\
=& \left[\mathcal{\hat G}_\sigma^{-1}-\hat J -
{\hat\Sigma}_\sigma\right]^{-1}(\vec r_i,t;\vec r_{i'},t'),
\end{split}
\label{eq:Glatt}
\end{equation}
i.e.~
\begin{equation}
G^\text{(loc)}_{i,\sigma}(t,t') = G^\text{(latt)}_{\sigma}(\vec r_i,t;\vec r_i,t').
\end{equation}
Here, the hatted quantities $\mathcal{\hat G}_\sigma$, $\hat J$, and ${\hat\Sigma}_\sigma$ denote the non-interacting Green's function,
the hopping, and the self-energy as operators acting on both, time and space coordinates.

\subsection{Observables}

We comment now on the measurement of some important physical observables within our method.
The \emph{spin-$\sigma$ occupancy} of site $i$ may be evaluated by
\begin{equation}
\begin{split}
\langle n_{i\sigma} \rangle(t) =& -\imag \cdot (-\imag)\cdot \lim_{\delta \to 0^+} \langle T_\mathcal{C} c_{i\sigma}(t)c_{i\sigma}^\dagger(t+\delta)\rangle \\
=& -\imag\mathcal{G}^\text{(loc)}_{i\sigma}(t,t+ 0^+)
\end{split}
\end{equation}
from the local Green's function.

The \emph{kinetic energy contribution of spin $\sigma$ at lattice site $i$} can be deduced from the lattice Green's function
\begin{equation}
e^\text{kin}_{i\sigma}(t) := +\imag \cdot \sum_{j\text{ is NN of } i} J_{ij,\sigma}(t) G^\text{(latt)}_\sigma(\vec r_j,t;\vec r_i,t).
\end{equation}
It is thus required to measure the equal-time hopping Green's functions from site $i$ to its nearest neighbors $j$.

A quantity of particular interest is the \emph{double occupancy $D_i(t)=\langle n_{i\uparrow}n_{i\downarrow}\rangle(t)$} as a function of time $t$ and lattice site $i$.
We can derive it from some elements of the lattice Green's function and its equal-time derivative, using the following relation 
(see Appendix \ref{app:doccformula}):
\begin{equation}
\begin{split}
\left.\frac{\partial}{\partial t} G^\text{(loc)}_{i,\sigma}(t,t')\right|_{t'=t^+}
\,=&\, U_i(t) D_i(t)\,\, \\
&\,+ \epsilon_{i}(t)\langle n_{i\sigma}\rangle(t) + e^\text{kin}_{i\sigma}(t).
\end{split}
\label{eq:docc}
\end{equation}

\section{Algorithm}
\label{sec:algo}
In this section, we outline the massively parallel algorithm required to solve Eqs.~\eqref{eq:defnSigma}, \eqref{eq:defnSigmatilde}, and 
\eqref{eq:Glatt} iteratively on a supercomputer.

\begin{figure}
\includegraphics[width=0.5\linewidth]{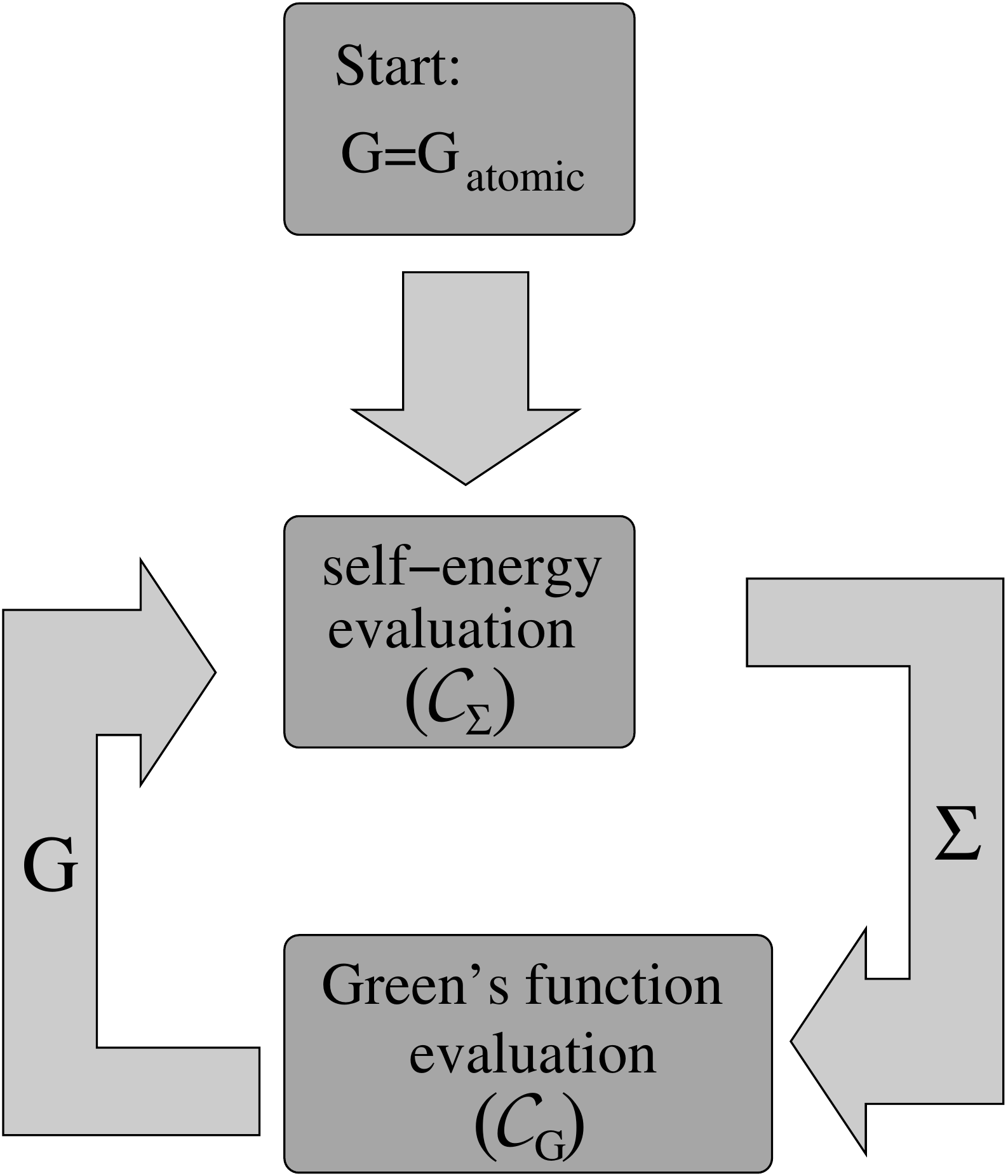}
\caption{The self-consistent strong-coupling algorithm as a flow chart. The configurations $\mathcal{C}_\Sigma$ and $\mathcal{C}_G$ are introduced in
Sec.~\ref{sec:algo} and correspond to evaluating Eqs.~\eqref{eq:defnSigma} and \eqref{eq:defnSigmatilde} (both $\mathcal{C}_\Sigma$), and 
Eq.~\eqref{eq:Glatt} ($\mathcal{C}_G$).}
\label{fig:algorithm}
\end{figure}

\subsection{Representation of the Green's functions and self-energies}
In order to represent the contour-ordered Green's function on a computer, the Kadanoff-Baym-Keldysh contour is 
discretized to $N_t$ time slices, as shown in Fig.~\ref{fig:Keldysh}. 
In the implementation used here, we chose $N_t=2N_{t,\text{real}} + N_{t,\text{imag}}$, where
$N_{t,\text{imag}} = N_t/32$. Here, $N_{t,\text{real}}$ is the number of time steps on each real
branch of the Kadanoff-Baym-Keldysh contour and $N_{t,\text{imag}}$ the number of time steps on the imaginary
time axis.
Furthermore, the system consists of a finite number $N_r$ of lattice sites.
Assuming that spatial symmetries such as reflection and rotation symmetries are given for not only the Hamiltonian but also the quantum-statistical states,
we can reduce the actual number of lattice sites $N_r$ within the algorithm by symmetry maps to the number $\tilde N_r$ which is the
number of sites in the irreducible wedge of the lattice.
Many sites can then be represented by the equivalence class with respect to the symmetry.
Note that the full number of lattice sites still plays a role in computational complexity when the propagation of excitations is considered. 
Since Dyson's equation performs an infinite resummation of such processes, it is relevant for the calculation of the Green's function. 
Appendix \ref{app:symmetries} describes the exploitation of symmetries in more detail.

\subsection{Global Layout} %
Computationally, the self-consistency condition in Eq.~\eqref{eq:selfcons} is solved iteratively using an alternating sequence of self-energy  
and Green's function evaluations. The self-energy evaluations are performed via Eqs.~\eqref{eq:defnSigma} and \eqref{eq:defnSigmatilde} and will
be described in detail in Sec.~\ref{subsec:selfenergyeval}. The
Green's functions required for self-consistency and measurements of observables are evaluated via Eq.~\eqref{eq:Glatt}
which will be described in Sec.~\ref{subsec:GreensFunctionEvaluation}.

Due to the structure of Eq.~\eqref{eq:defnSigmatilde}, the computation of $\Sigma_i(t,t')$ at lattice site $i$ requires only the Green's functions of neighboring lattice sites. 
However, the second step, i.e. the Dyson's equation evaluation in Eq.~\eqref{eq:Glatt} of the lattice Green's requires self-energy information from all lattice sites.
From a computational perspective, these demands are quite different in terms of optimal memory arrangement and distribution of tasks over a large set of compute nodes. As a consequence, we 
define two different configurations of the simulation. 
\begin{figure}
\centering
\includegraphics[width=0.9\linewidth]{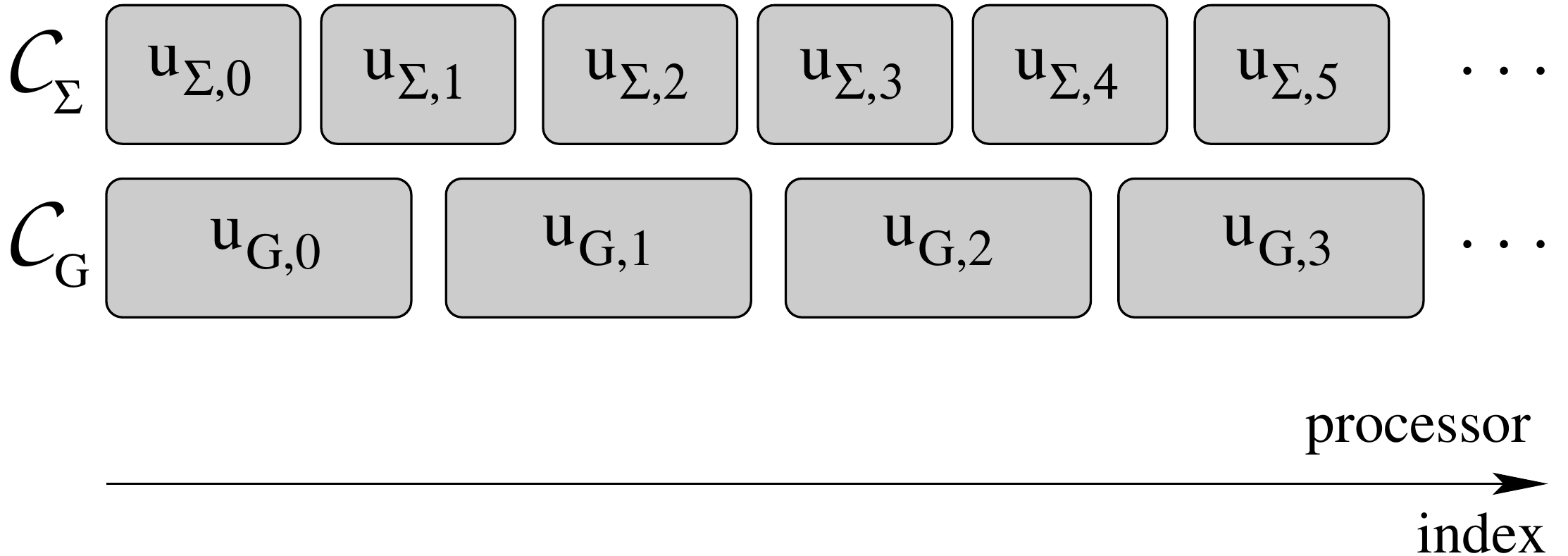}
\caption{Global configurations for the parallelization scheme.}
\label{fig:processorgroups}
\end{figure}
The Green's function evaluation in Eq.~\eqref{eq:Glatt} is performed in what we call the $\mathcal{C}_G$ configuration of the simulation.
The self-energy computation is done in the $\mathcal{C}_\Sigma$ configuration. Fig.~\ref{fig:algorithm} shows a schematic flowchart for the
computation.

The two different global machine states are sketched in Fig.~\ref{fig:processorgroups}. All processors are thought to be arranged along
the direction of the abscissa in Fig.~\ref{fig:processorgroups}. Each configuration defines groups of processors which share tasks,
we call them self-energy units $u_{\Sigma,i}$ and Green's function units $u_{G,i}$. 
The self-energy unit $u_{\Sigma,i}$ evaluates self-energies for the $i$-th range of representative sites. The 
Green's function units evaluate relevant information for the $i$-th range of symmetry-representative site indices for blocked rows of the lattice 
Green's function.
There is an optimal value for the size of the units 
which is usually different for $u_{\Sigma,i}$ and $u_{G,i}$. The determining factors for the optimal size will be elaborated below.

\subsection{Self-energy evaluation for unit $u_{\Sigma,i}$}
\label{subsec:selfenergyeval}
By a given self-energy unit $u_{\Sigma,i}$, the self-energy $\Sigma_n(t,t')$ is computed for a certain range of 
(representatives of) sites $n$ for all $t,t'$ using Eqs.~\eqref{eq:defnSigma}
and \eqref{eq:defnSigmatilde}. The expressions involve an on-the-fly evaluation of the cumulant $\tilde{\mathcal G}^{II}$ using the 
tabulated values of Eq.~\eqref{eq:cumulanttabulation}, as well as local Green's functions which
do not require any permanent storage on $u_{\Sigma,i}$.
Memory-wise, the unit is required to have access to the local Green's functions of neighboring sites, as well as to the value of the self-energy from the
previous iteration. The latter is necessary, because in our implementation, Eq.~\eqref{eq:defnSigma} is regularized 
as follows:
\begin{equation}
\hat \Sigma = \lambda_\text{mix}\hat \Sigma_\text{new} + (1-\lambda_\text{mix}) \hat \Sigma_\text{old},
\label{eq:Sigmamix}
\end{equation}
where $\lambda_\text{mix} \approx 0.7$ is a linear mixing parameter which stabilizes the convergence by averaging between the current and the previous iteration. 
Convergence is reached once
\begin{equation}
N_\text{bd}^{-1}\|{\hat \Sigma}_\text{new} - {\hat\Sigma}_\text{old}\|_2 + \|{\hat \Sigma}_\text{new} - {\hat\Sigma}_\text{old}\|_\infty \leq \delta_\text{sc},
\end{equation}
where $\|\cdot\|_2$ is the Frobenius norm of the self-energy matrix with $N_\text{bd}$ entries on the block diagonal, and 
$\|\cdot\|_\infty$ is element-wise maximum norm of the matrix. A reasonable value for the accuracy of the self-consistency condition is
\begin{equation}
\delta_\text{sc} \approx 10^{-8} U_0,
\end{equation}
where $U_0$ is some typical value of the interaction strength (which often serves as the energy unit). Convergence is usually reached after approximately ten iterations.

In all relevant cases we have encountered so far, the self-energy evaluation step is computationally more costly than the Green's function evaluation, if the 
latter has been optimized appropriately. This is due to the fact, that the contraction of cumulant indices in Eq.~\eqref{eq:defnSigmatilde} scales with the fourth power of $N_t$.
The evaluation of the cumulant contraction is however massively parallelizable, i.e. no significant communication even within the $u_{\Sigma,i}$ is required during 
the computation. 

Thus, the size of the cumulant units can be chosen rather freely regarding the aspect of communication between CPUs within the unit. A small size is preferable, due to the 
typically small memory consumption. For convenience, when switching between configurations $\mathcal{C}_\Sigma$ and $\mathcal{C}_G$, the self-energies just remain 
within the compute nodes of the respective processors of the self-energy units.
However, it is better to ensure that each self-energy unit deals with more than one lattice site, because we encounter the situation that the time to evaluate the 
self-energy varies substantially from lattice site to lattice site. If the site indices are assigned randomly, this effect averages out for a sufficient number of sites.
As a rule of thumb, it is reasonable to assign self-energies for sixteen random lattice sites to each $u_{\Sigma,i}$.

\subsection{Green's Function evaluation for unit $u_{G,i}$}
\label{subsec:GreensFunctionEvaluation}
Let us have a closer look at the Dyson's equation in Eq.~\eqref{eq:Glatt}. It
contains a block-diagonal part 
\begin{equation}
\mathcal{\hat G}_\sigma^{-1}-{\hat\Sigma}_\sigma =
\text{diag}_{i\,\in\,\text{lattice}}\,(\mathcal{G}_{i,\sigma}^{-1}-{\Sigma}_{i,\sigma}),
\label{eq:blockdiagdyson}
\end{equation}
and a sparse off-diagonal part given by the hopping matrix $\hat J$. The
latter is composed of a contour-$\delta$-function in time and a
tight-binding structure in real space. In particular, $\hat J$ is a small
quantity, due to the very nature of the strong-coupling expansion.
This fact is exploited algorithmically.

It is insightful to change notation in Eq.~\eqref{eq:Glatt}. We can
write the lattice Green's function in a Dirac type notation, as a matrix
element
\begin{equation}
G^\text{(latt)}_\sigma(\vec r,t;\vec r',t') = \langle \vec r,t
|\left(\frac{1}{\mathcal{\hat G}_\sigma^{-1}-\hat J -
{\hat\Sigma}_\sigma}\right)|\vec r ',t'\rangle,
\label{eq:Glatt2}
\end{equation}
where $\{|\vec r,t\rangle\}_{\vec r,t}$ is an orthonormal basis associated with space and time variables.
Each $u_{G,i}$ computes $G^\text{(latt)}_\sigma(\vec r,t;\vec r',t')$ at the sites $\vec r$ assigned to it
for all $\vec r', t,t'$ and stores the local and hopping Green's functions as required.
Determining the lattice Green's function corresponds to solving $N_r\cdot N_t$
linear equations in $N_r\cdot N_t$ variables. That is, the $(N_r N_t)^2$ matrix elements of the Green's function
are determined by solving such linear equations $N_r N_t$ times.
Typically, in our application, $N_t$ is at least 512 or 1024 and $N_r$ is around 10000.
Note that since the simulation discretizes the points on the Kadanoff-Baym-Keldysh contour with $N_t$ time
slices, there is a finite size of time slices $\Delta t$. 
The effect of a finite $\Delta t$ usually affects the accuracy of the calculations, and thus we perform
a quadratic extrapolation of the simulation results from three finite values of $\Delta t$ to $\Delta t\to 0$.
Cross-checks with linear extrapolations show the quadratic extrapolation is superior in most instances.

The matrix to be inverted in Eq.~\eqref{eq:Glatt2}, 
$\left(\hat G^\text{(latt)}_\sigma\right)^{-1}$, is typically around $5\cdot 10^{6}\times 5\cdot 10^{6}$
dimensional. 
Its block structure can be written as
\begin{equation}
\begin{split}
&\left(\hat G^\text{(latt)}_\sigma\right)^{-1}= \\
&\begin{pmatrix}
B_1    & 0   & \cdots & 0      & -J_{1,NN(1)_i}     & 0      &  \cdots & 0 \\
0      & B_2 & 0      & \cdots & -J_{2,NN(2)_i}     & 0      & \cdots \\
\vdots & 0   & B_3    & 0      & \cdots \\
       &     &        & \ddots \\
       &     &        &        &  \\
       &     &        &        &                    &  \\
       &     &        &        &                    &        &         & \vdots \\
       &     &        &        &                    &        & \ddots  & 0\\
       &     &        & \cdots & -J_{N_r,NN(N_r)_i} & \cdots &   0     & B_{N_r}
\end{pmatrix},
\end{split}
\label{eq:Ginvblockstruct}
\end{equation}
where each block represents an $N_t\times N_t$ matrix and 
\begin{equation}
B_i = \mathcal{G}_{i,\sigma}^{-1} - \Sigma_{i,\sigma},
\label{eq:blockdiagdysonelement}
\end{equation}
as given in Eq.~\eqref{eq:blockdiagdyson}.
Each row and each column in Eq.~\eqref{eq:Ginvblockstruct} contains at most 
4 hopping matrices $J_{n,NN(n)_i}$ to the nearest neighbors $NN(n)_i$, $i=1,\dots, 4$ (in two dimensions).
The hopping matrices $J_{n,NN(n)_i}$ are essentially
diagonal matrices whose diagonal elements are the time-dependent hopping amplitudes.
The matrix in Eq.~\eqref{eq:Ginvblockstruct} is extremely sparse, while its inverse 
$\hat G^\text{(latt)}_\sigma$ is dense. However, due to the small numerical value of the hopping
in the strong-coupling expansion, matrix elements of $\hat G^\text{(latt)}_\sigma$
fall off as a function of spatial distance and eventually become irrelevant;
in other words, the lattice Green's function is typically block diagonal dominant.
This fact can be exploited in a numerically controlled way by utilizing an 
iterative procedure. We choose the Generalized Minimal Residue (GMRES) method as 
a solver \cite{GMRES}.

The GMRES method considers an equation system
\begin{equation}
y = \hat Ax,
\end{equation}
as well as an "almost" correct solution
\begin{equation}
\tilde x = \mathcal{P}(y),
\end{equation}
where $\mathcal{P}$ is the  so-called \emph{preconditioner}. 
$\mathcal{P}$ is an arbitrary operator whose action on $y$ is cheap to compute but is
a good approximation to $\hat A^{-1}y$. In our case, due to the small numerical value of the 
hopping, a natural choice for the preconditioner is the inverse block diagonal matrix
\begin{equation}
\mathcal{P}(y) := {\hat B}^{-1} y,
\end{equation}
where
\begin{equation}
\hat B := \mathcal{\hat G}_\sigma^{-1} - {\hat\Sigma}_\sigma
\label{eq:definitionB}
\end{equation}
as defined earlier in Eqs.~\eqref{eq:blockdiagdyson} and \eqref{eq:blockdiagdysonelement}.
Further details of the GMRES method as applied to the Dyson's equation in the inhomogeneous strong-coupling expansion are 
discussed in Appendix \ref{app:gmres}.

We distribute the task of applying GMRES to each unit vector $|e_j\rangle$ of the vector space $\mathbb{C}^{N_t N_r}$
within the set of Green's function units $\{u_{G,i}\}_{i=1,\dots,N_{G}}$. Each Green's function unit
must store all blocks of the block diagonal $\hat B$. Memory allocated to each CPU within a unit thus defines a 
lower bound to the size of a Green's function unit through this requirement in the following way:
\begin{equation}
|u_{G,i}| \geq \left\lceil\frac{\tilde N_r \cdot N_t^2 \cdot 16\,\text{Bytes per complex}}{\text{memory per CPU}}\right\rceil,
\end{equation}
assuming a double precision representation of complex numbers. If one also decides to store the
preconditioner, this value doubles.
\begin{figure}
\centering
\includegraphics[width=0.8\linewidth]{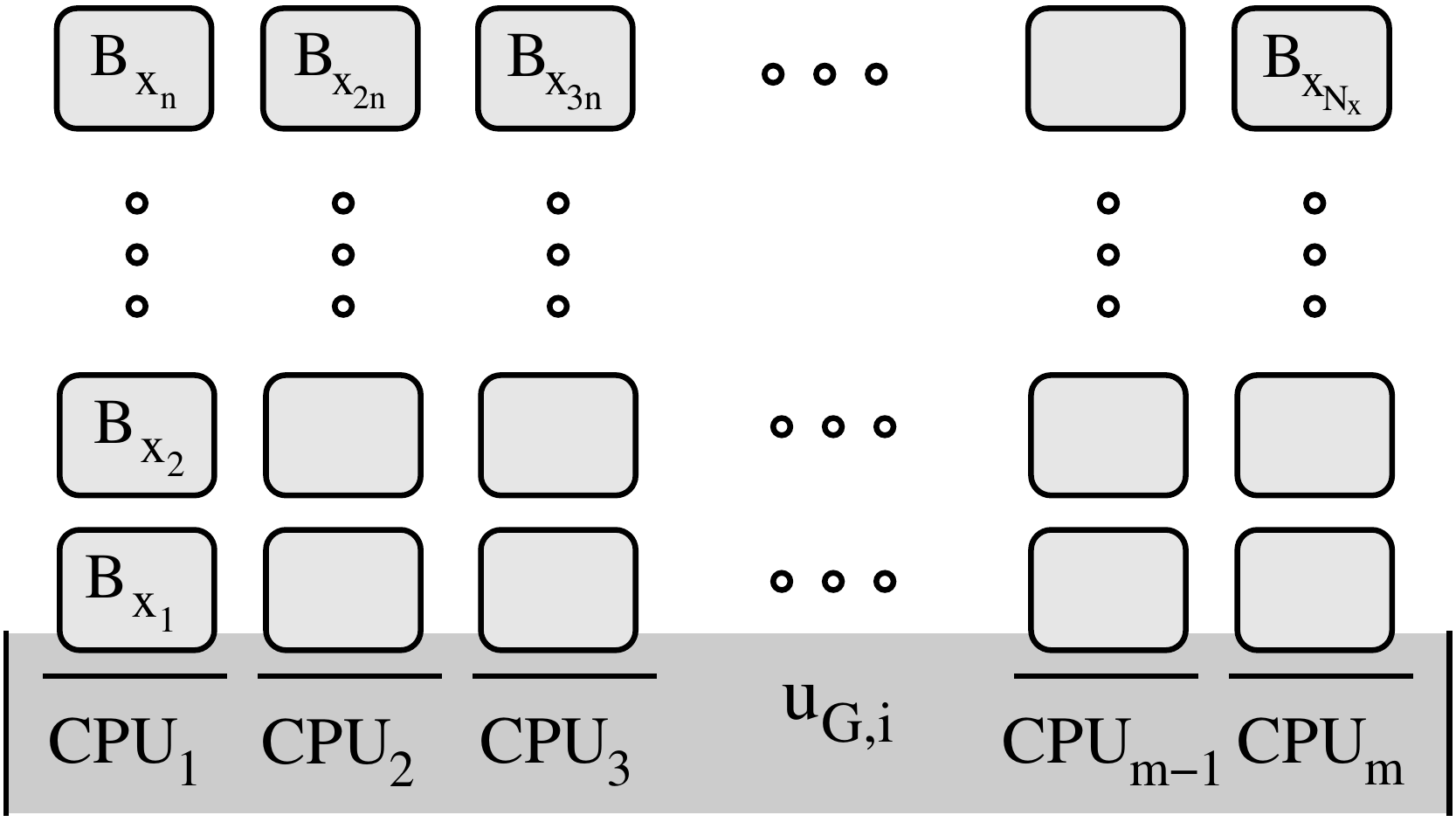}
\caption{Distribution of the non-zero entries (lightly shaded boxes) $B_{x_\nu}(t,t')$ of the block-diagonal matrix $\hat B$ within each unit $u_{G,i}$ 
(darkly shaded area). If symmetries apply, only the blocks associated with representatives of lattice sites need to be stored, and $N_r$ 
is then taken as $\tilde N_r$.}
\label{fig:storageB}
\end{figure}
We store the blocks of $\hat B$ as depicted in Fig.~\ref{fig:storageB}.

Practically, within each $u_{G,i}$, an efficient application of $\hat B$ and $\hat B^{-1}$ to a given vector $|x\rangle$ has to 
be achieved. The performance of these operations crucially relies on the small size of $u_{G,i}$ and on
a good load-balancing within $u_{G,i}$. The former is due to an increased number of communication events through relatively
slow connections between CPUs and also due to its role in the latter.
In order to avoid an unnecessarily large number of CPUs in $u_{G,i}$, a minimization of memory consumption plays a key role.
We construct the representation of $\hat B$ within a Green's function unit by distributing respective self-energies
to the CPUs dealing with specific parts of $\hat B$.
A practical example is shown in table \ref{tab:para}.
\begin{table}
\begin{tabular}{c|c|c|c|c}
$N_t$ & CPUs (walltime) & size of $u_{\Sigma,i}$ & size of $u_{G,i}$ & $n_b$ \\
\hline
256   &   16  (24h)     & 16                     & 2                 & 256 \\
512   &   512 (24h)     & 16                     & 4                 & 512 \\
1024  &  8192 (50h)     & 256                    & 64                & 256
\end{tabular}
\caption{Table of typical parallelization parameters on a Cray XE6 (2GB memory per CPU) for the application presented in Sec.~\ref{subsec:numresults}.
The number $n_b=N_rN_t/N_b$ denotes the block size in the parallel GMRES implementation.
}
\label{tab:para}
\end{table}

A further important optimization when applying GMRES takes advantage of both cache optimizations in Basic 
Linear Algebra Subprograms (BLAS)\cite{BLASLAPACK} level 3
routines and a reduction of network latency effects --- rather than letting $u_{G,i}$ apply GMRES individually to each 
unit vector $|e_j\rangle$ of its concern serially, we choose a parallel implementation. It applies the algorithm to blocks rather 
than vectors. Thus, highly optimized BLAS level 3 rather than multiple calls of BLAS level 2 are used. In addition, less 
communication processes between CPUs occur.
More specifically, we can write this blockwise procedure as solving the blocked equation system
\begin{equation}
\begin{pmatrix}
\hat E_1 & \hat E_2 & \cdots & \hat E_{N_b}
\end{pmatrix}
= (\hat B-\hat J) \cdot
\begin{pmatrix}
\hat \Gamma_1 & \hat \Gamma_2 & \cdots & \hat \Gamma_{N_b}
\end{pmatrix},
\label{eq:blockedGMRESproblem}
\end{equation}
where 
\begin{equation}
\hat E_i = \sum_{j=(i-1) N_r N_t / N_b}^{i N_r N_t / N_b} |e_j\rangle \langle e_j|,
\label{eq:unitblock}
\end{equation}
with appropriately defined unit vectors $|e_j\rangle$ which belong to a single lattice site in the notation introduced above 
with Eq.~\eqref{eq:Glatt2}. The blocks $\hat \Gamma_i$ are the respective blocks of $\hat G_\sigma^\text{latt}$.
Typical block sizes are listed in Table \ref{tab:para}.

In the GMRES procedure, this definition of blocked equations which are 
solved one after another has the following advantage.
The method operates iteratively, where the start value is the preconditioner solution, which is still localized at some 
lattice site: $\hat B^{-1} |e_j\rangle$ (or $\hat B^{-1} \hat E_i$).
The computation of $\hat B^{-1}$ is done with LAPACK calls\cite{BLASLAPACK} for the diagonal blocks by the CPUs assigned to them according to
Fig.~\ref{fig:storageB}.
By means of the spatial structure of this then iteratively refined approximation to the solution, only the application of $\hat B - \hat J$
introduces further lattice sites to the problem.
As a consequence, if GMRES converges quickly for a given numerical accuracy of the GMRES method, non-zero elements only occur within the spatial 
vicinity associated to the considered $\hat E_i$ block in the blocked 
representation of the converging GMRES solution 
$\hat X =\begin{pmatrix} x_1 & x_2 & \cdots & x_{N_rN_t/N_b} \end{pmatrix} $ (see App.~\ref{app:gmres} for details). Indeed, the GMRES procedure
converges very rapidly, because the hopping $\hat J$ between adjacent lattice sites is required to be small in the strong-coupling expansion.

For the block representations $\hat X$ of iteratively refined GMRES solutions, we thus choose to only store non-zero subblocks and distribute 
them within the 
$u_{G,i}$ units using the storage scheme for $\hat B$ (see above and Fig.~\ref{fig:storageB}). In the case that $\hat X$ contains non-zero contributions for reducible lattice sites 
(this happens when $\hat E_i$ is located near the boundary of the range of site representatives, see Appendix \ref{app:symmetries}),
the respective block is handled by the CPU associated to its equivalence class.
This minimizes both communication and memory consumption. The former is because applying the preconditioner is only a local operation and applying $\hat A$ requires
only communications with units storing neighbors of the respective lattice sites.
The size of the blocks in $\hat X$ is to be chosen as large as possible. However,
memory in $u_{G,i}$ is usually very limited, so a tradeoff in unit size and
block length has to be made.

Let us also comment on reasonable values for the GMRES convergence parameter. The GMRES method for our matrix inversion
runs until a certain accuracy for the result is achieved, i.e.
\begin{equation}
\|e_j - \hat A x\|_2 \leq \delta_\text{GMRES},
\label{eq:GMRESconvergence}
\end{equation}
where $\delta_\text{GMRES}$ is the desired numerical precision and $\|\cdot\|_2$ is the 
Euclidian norm.
For all practical purposes we encountered so far, a value 
\begin{equation}
\delta_\text{GMRES} = 10^{-2}
\label{eq:GMRESconvergence2}
\end{equation} 
has been sufficient.
This surprisingly large value was verified by comparing to simulations with higher accuracy, that is $\delta_\text{GMRES} = 10^{-3}$, for the physical
systems studied in this paper at several parameter values. The plots of numerical results of interest are identical to the eye. Similar 
tests were done for completely homogeneous systems by comparing to a numerically exact implementation in momentum space.
It may be that for different applications than the one presented here a smaller value of $\delta_\text{GMRES}$ is required.
In order to understand the meaning of $\delta_\text{GMRES}$ better it may be useful to compare it to the dimension of the vector it constrains. In our case,
the dimension of $e_j - \hat A x$ is $N_r N_t \geq 1\times 10^6$. Thus, if one chooses to normalize the convergence criterion in Eqs.~
\eqref{eq:GMRESconvergence} and \eqref{eq:GMRESconvergence2}
by the dimension, the constraint reads $10^{-8}$.
In this context it may also be worthwhile to consider the fact that the GMRES procedure only involves transformations with $\hat B^{-1}$,
$\hat B$, and $\hat J$. That is, it applies only transformations which comply with the causal structure of the Green's function and do not
introduce artificial discontinuities with respect to the time variables in the end result for the Green's function which are in principle 
part of the vector space which is being searched by the algorithm. In other words, the physical choice of the preconditioner already 
constrains the solution space so drastically that even a relatively large value of $\delta_\text{GMRES}$ might be sufficient.

With all these optimizations, the Green's function evaluation typically consumes no more than five to ten percent of the
time required for the self-energy evaluation on a Cray XE6 machine in the application to lattice depth modulation spectroscopy
described below. The optimizations are necessary to speed up the Green's function evaluation appropriately, because we encountered 
increases in speed by a factor of at least 10 and up to 1000, for 
each blockwise application of GMRES, distributed storage of the GMRES vector blocks, and random assignment of site indices.
In other applications than lattice depth modulation, with a large value of the hopping applied for a longer period of time, the 
requirement of computer time for the Green's function evaluation may still exceed the time to evaluate the self-energy. However, we 
find these requirements to be within reasonable bounds.

\subsection{Switching between the global configurations}
\label{subsec:confswitch}
The only time when global communication and synchronization across all processors is required is when either the self-energy or the
Green's function evaluations are finished. Then, convergence has to be checked, and the global configurations $\mathcal{C}_\Sigma$ and
$\mathcal{C}_G$ have to be replaced by each other. This requires point-to-point communications of individual processors across
the machine and broadcasts within smaller groups of processors which all require the same data. 
\begin{figure}
\subfloat[]{
\includegraphics[width=\linewidth]{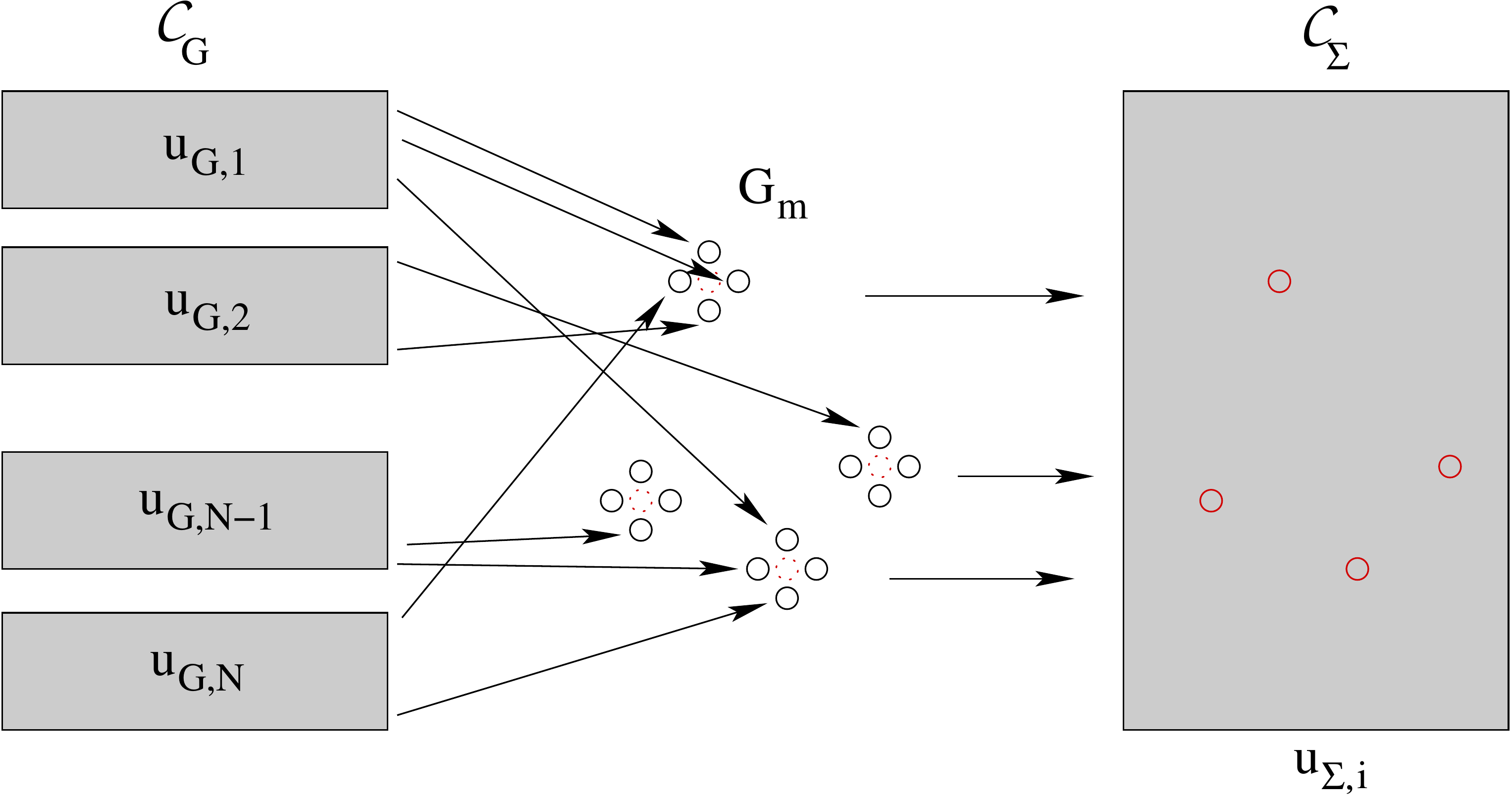}
\label{subfig:CGtoCS}
}
\\
\subfloat[]{
\includegraphics[width=\linewidth]{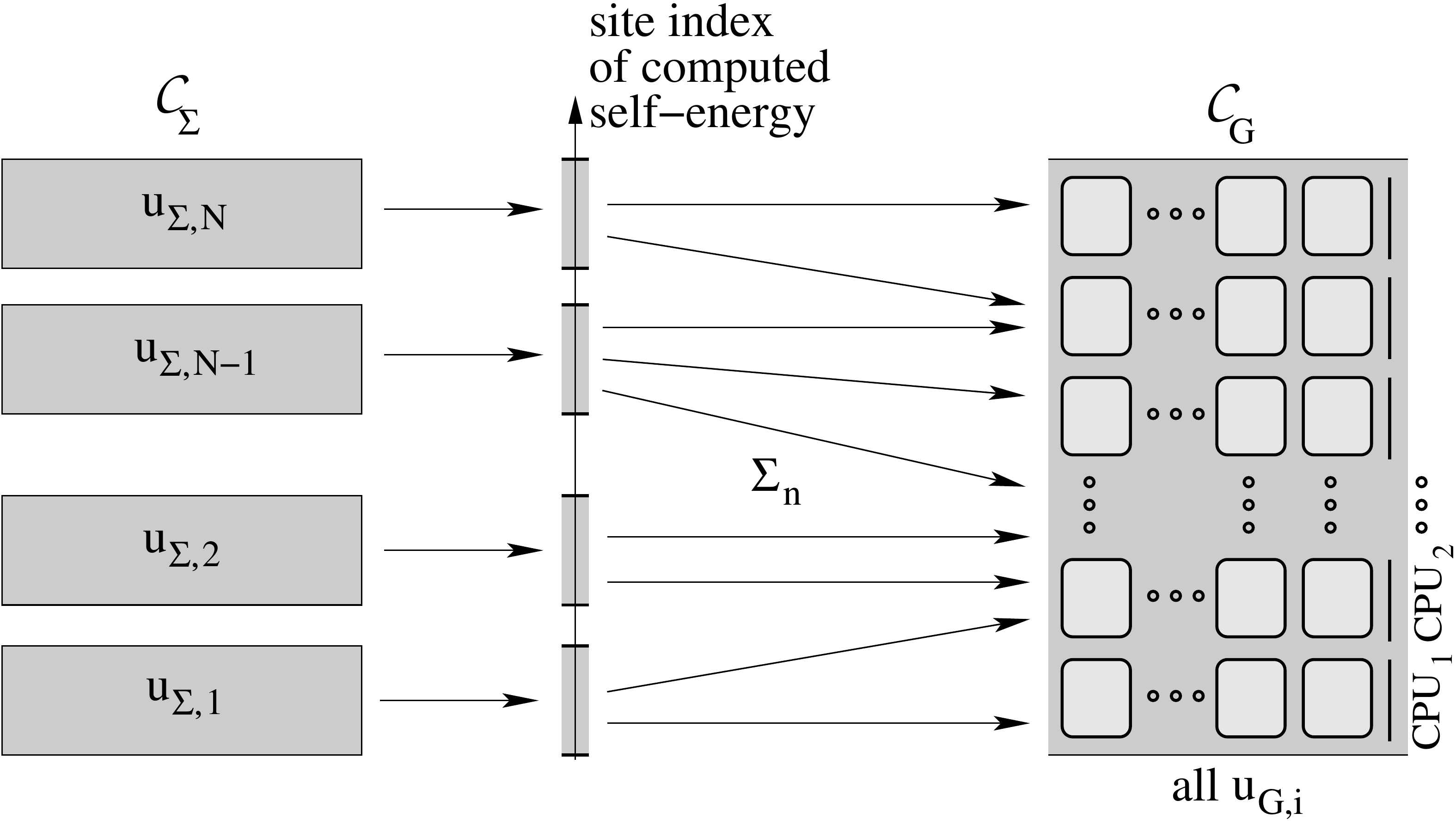}
\label{subfig:CStoCG}
}
\caption{(color online) Global data transfers during the configuration switches $\mathcal{C}_\Sigma \leftrightarrow \mathcal{C}_G$.
(a): Transfers from the $u_{G,i}$ units to a given $u_{\Sigma,i}$ unit during the configuration 
switch $\mathcal{C}_G \to \mathcal{C}_\Sigma$. Only the local Green's functions $G_m(t,t')$ on the nearest neighbors of the assigned self-energies $\Sigma_n(t,t')$ are required by each $u_{\Sigma,i}$ due to Eq.~\eqref{eq:defnSigmatilde}.
(b): Transfers of the full self-energy data $\Sigma_n(t,t')$ from the $u_{\Sigma,i}$ units to all $u_{G,i}$ units.
The process can be improved by sending the data only to $u_{G,1}$ and then broadcasting the data internally to the set of 
equivalent member processors $\text{CPU}_j$, $j=\text{const.}$, of each $u_{G,i}$, $i>1$.
The memory structure of the Green's function unit is identical to the block-diagonal storage scheme introduced in Fig.~\ref{fig:storageB}.
}
\label{fig:globalconfigswitches}
\end{figure}
The occuring communication events are displayed in Fig.~\ref{fig:globalconfigswitches}.
During the switch $\mathcal{C}_G\to\mathcal{C}_\Sigma$ (Fig.~\ref{subfig:CGtoCS}), the Green's function contents on the nearest neighbors of the sites 
assigned to a given $u_{\Sigma,i}$ have to be sent from the Green's function unit which computed them.
Due to the possibly random assignment of site indices $m$ to spatial coordinates 
${\vec r}_m$, the input to $u_{\Sigma,i}$ is collected from various $u_{G,j}$. The
storage scheme used for the results within $u_{G,j}$ determines the actual processors
which send the data.

When switching from $\mathcal{C}_\Sigma$ to $\mathcal{C}_G$, spatial ranges tasked to specific parts of each Green's function unit have to
be sent from the self-energy units which contain this information (Fig.~\ref{subfig:CStoCG}).
The storage pattern for $u_{G,i}$ is already the one specified for the block diagonal matrix denominator of the Dyson's equation (Fig.~\ref{fig:storageB}).
To finish the change of configurations, an in-place substitution of the self-energies by the respective blocks $B_{x_i}$ is performed by computing the 
respective atomic Green's functions and inserting $B_{x_i} = \mathcal{G}_i^{-1} - \Sigma_i$, as defined in Eq.~\eqref{eq:definitionB}.
The process $\mathcal{C}_\Sigma\to\mathcal{C}_G$ is typically more time-consuming 
than $\mathcal{C}_G\to\mathcal{C}_\Sigma$. 
However, it can be optimized by using broadcasts between groups of processors with the same data
requirements: each $j$-th CPU in $u_{G,i}$ requires the same data set to operate.

The switching processes cost no more than five percent of the total computation time and are thus negligible. However, the implementation involves a considerable amount of 
book-keeping.

\subsection{Summary and notes on the implementation}
Let us summarize the algorithm and also provide some implementation details on the way.
As the method implements the self-consistent solution of Eq.~\eqref{eq:selfcons}, the algorithm is split in two steps: the self-energy
evaluation (Eqs.~\eqref{eq:defnSigma} and \eqref{eq:defnSigmatilde})  and the Green's function evaluation (Eq.~\eqref{eq:Glatt}). 
Since the computation of the self-energy and the evaluation of the Green's function have
different requirements in terms of computational resources on the supercomputer, they use different data structures and 
collaboration patterns amongst the CPUs.
We refer to the data structures of the algorithm in the self-energy evaluation state as the configuration $\mathcal{C}_\Sigma$ and to the data structures of the 
algorithm in the Green's function evaluation state as the configuration $\mathcal{C}_G$.
The respective configurations are subdivided into mutually independent units $u_{\Sigma,i}$ and $u_{G,i}$ spanning several CPUs and the
memory associated with them, respectively. This approach is tailored to a cluster, rather than a shared-memory architecture.

If one chooses to employ the Message Passing Interface (MPI) standard in order to implement the algorithm, it is advantageous to use the \texttt{MPI\_Group} 
feature to ensure an efficient communication within the units \cite{mpibook}.
It turns out to be useful to define internal communicators for the  $u_{G,i}$ and $u_{\Sigma,i}$, respectively, as well as for sets of processors with shared requirements, such
as the $n$-th processor of each $u_{G,i}$, since they all require the same self-energies. Within these shared-interest communicators, data can be broadcasted efficiently.
It may also be reasonable to use advanced MPI features to perform an optimization with respect to the network topology of 
the supercomputer, such that communication within the $u_{G,i}$ units is optimal and equally fast for all $i$. 
In contrast, the communication within $u_{\Sigma,i}$ is not time-critical, because the main effort, computing the integrals in Eq.~\eqref{eq:defnSigmatilde}, is done by 
each processor in $u_{\Sigma,i}$ independently.

Let us now comment on the implementation of Eq.~\eqref{eq:defnSigmatilde}. Each $u_{\Sigma,i}$ computes $\tilde \Sigma$ for a certain range of sites. Here, the 
main effort is the contraction of the time indices. The atomic-limit cumulant Green's function $\mathcal{G}^{II}$ is computed on the fly using tabulated values of the exponentials in Eq.~\eqref{eq:cumulanttabulation}. 
The computational effort of Eq.~\eqref{eq:defnSigmatilde} scales with $N_t^4$ and is the computationally most costly operation. However, it may also be implemented on GPUs, due 
to little memory and bandwidth requirements. 
After having evaluated Eq.~\eqref{eq:defnSigmatilde}, the units $u_{\Sigma,i}$ compute the updated self-energy using Eqs.~\eqref{eq:defnSigma} and \eqref{eq:Sigmamix}. Then, as 
described in section \ref{subsec:confswitch}, the resulting local self-energies are sent to the $u_{G,i}$ units which may or may not overlap with the respective $u_{\Sigma,i}$. 
Within each $u_{G,i}$, the self-energy for all sites has to be available and is thus equally distributed over the CPUs according to the storage pattern depicted
in Fig.~\ref{fig:storageB}. It is advised to keep the self-energy results in $u_{\Sigma,i}$ for the next update as described by Eq.~\eqref{eq:Sigmamix}, even though the
machine changes to configuration $\mathcal{C}_G$ in the meantime. This is because the $\hat \Sigma_\text{old}$ in Eq.~\eqref{eq:Sigmamix} has to be available in the self-energy computation
of the next iteration of the self-consistency loop. In order to minimize the memory consumption of the relevant range of $\hat \Sigma_\text{old}$, it can be distributed equally within each $u_{\Sigma,i}$.

In order to establish the configuration $\mathcal{C}_G$ to compute Eq.~\eqref{eq:Glatt}, the self-energies in $u_{G,i}$ are then replaced by the blocks of the matrix $\hat B$ according to 
Eq.~\eqref{eq:definitionB}. Optionally, the elements of the preconditioner $\hat B^{-1}$ can also be computed and stored at this point in time, also using the previous storage pattern. 
However, this competes with the requirement to keep $u_{G,i}$ small, because the action of the preconditioner can also be computed on the fly from $\hat B$ with a smaller memory requirement.

Having fully set up the $\mathcal{C}_G$ configuration, Eq.~\eqref{eq:Glatt} is written as the vectorized linear Eq.~\eqref{eq:blockedGMRESproblem} as described in 
section \ref{subsec:GreensFunctionEvaluation}.
The key variable is the bundle of GMRES vectors $\hat X$ whose initial value $\hat X_0$ is the preconditioner $\mathcal{P}=\hat B^{-1}$ applied to a bundle of unit vectors, as in Eq.~\eqref{eq:unitblock}. 
$\hat X_0$ has only non-zero entries at a single site index.
$\hat X$ is also stored according to the scheme in Fig.~\ref{fig:storageB}. A good optimization here is to store only nonzero components of $\hat X$ emerging from $\hat X_0$ due to the application of 
the hopping matrix. For this purpose, each processor in $u_{G,i}$ can keep track of the sites with non-zero elements in $\hat X$ based on the site index of
$\hat X_0$ and the number of hopping events applied to $\hat X$ during the GMRES procedure elaborated in App.~\ref{app:gmres}. Once a hopping occurs
due to the application of $\hat B-\hat J$, a given processor may have contributions to be stored and/or added to a value assigned to another processor within the storage scheme
(compare to the block structure of $\hat B -\hat J$ in Eq.~\eqref{eq:blockedGMRESproblem}). Such
transmissions are the major communication events within $u_{G,i}$. The required communication bandwidth within $u_{G,i}$ can only be minimized by assigning connected spatial domains 
with a minimal surface to single processors within $u_{G,i}$. However, doing so is strongly disadvantageous in the case that the GMRES is converging very rapidly, i.e.~if the hopping
is small and $t_\text{max}$ is small. In this case, all but one of the CPUs in $u_{G,i}$ will remain idle, because the non-zero elements in $\hat X$ do not leave the CPU storing the non-zero 
elements of $\hat X_0$. Thus, for a rapidly converging GMRES, a random assignment of the site leading to largely scattered domains is more appropriate.

Once each $u_{G,i}$ has computed the Green's functions on the spatial range assigned to it, the different spatial components of $G_\text{loc}$ are distributed to the  
$u_{\Sigma,i}$ units which require them for evaluating the Green's function sum in Eq.~\eqref{eq:defnSigmatilde} in order to start the next iteration.

\section{Results}
\label{sec:results}
In this section, we present results of the algorithm for trapped atoms in a two-dimensional optical lattice which is subject to
a periodic modulation of the lattice depth. We compare this method to a homogeneous version of the algorithm which was previously 
successfully applied to a lattice depth modulation experiment within the LDA \cite{dirks2013}.

The LDA is generally expected to yield good results for systems without mass transport. This is a well-established observation in 
equilibrium \cite{Assmann2012}.
We show that also in a nonequilibrium scenario without mass transport, the high accuracy of the LDA can be explicitly demonstrated. 
At the same time we validate our direct computational approach.

\subsection{Lattice depth modulation}
In lattice-depth modulation spectroscopy, the atoms are subject to a time-dependent optical lattice potential 
\begin{equation}
V(\vec r, t) = V_\text{trap}(\vec r) + V_\text{lattice}(\vec r, t).
\end{equation}
The trap potential does not depend on time and has the parabolic shape
\begin{equation}
V_\text{trap}(\vec r) \propto |\vec r|^2.
\end{equation}
The lattice potential satisfies
\begin{equation}
V_\text{lattice}(\vec r, t) = V(t) \sum_{i=1}^2 \sin^2 (kx_i),
\end{equation}
which contains the time-dependent lattice depth
\begin{equation}
V(t) = V_0 + \chi_{[0, t_\text{mod}]} (t) \cdot \Delta V \sin \omega t.
\end{equation}
We assume that the lattice is modulated over a finite time interval $[0, t_\text{mod}]$ and that 
the system is in an initial thermal state at time $t=0$.
Numerically, we start the simulation at an earlier point in time, in order to be able to check for convergence, as discussed in Fig.~\ref{fig:Keldysh}.
The lattice constant $k=2\pi/\lambda$ is defined by the laser wavelength $\lambda$.
The single-particle Hamiltonian
\begin{equation}
H_\text{single} (t) = -\frac{\hbar^2}{2m} \vec\nabla^2 + V(\vec r, t)
\end{equation}
yields the recoil energy $E_R=\hbar k^2 / 2m$ as a natural choice for an energy unit. In order to compute the coefficients 
of the many-body Hamiltonian in Eq.~\eqref{eq:model} from the single-particle Hamiltonian, we insert the constant hoppings $J$ and interactions $U$
of a translationally invariant lattice. These can be computed easily with maximally localized Wannier functions \cite{Kohn1959}.

Due to the time-dependence of the lattice depth $V(t)$, we also obtain a time-dependent interaction $U(t)$ and hopping $J(t)$.
We write the initial values of the interaction and the hopping as $U_0$ and $J_0$, respectively.
In these units, the trap potential can be written as
\begin{equation}
V_\text{trap}(\vec r) = J_0 |\vec r / \rho_\text{trap}|^2.
\end{equation}
Hence, $\rho_\text{trap}$ can be interpreted as the length scale on which the trap potential reaches the strength of the initial hopping amplitude.
It is important to keep $\rho_\text{trap}$ larger than a couple of lattice spacings, since otherwise the trap potential interferes 
drastically with the hopping between neighboring sites and the density changes too fast for the LDA to be accurate.

\subsection{Numerical Results}
\label{subsec:numresults}
\begin{figure*}
\centering
{
\includegraphics[trim=20 150 20 510, clip, width=0.3\linewidth]{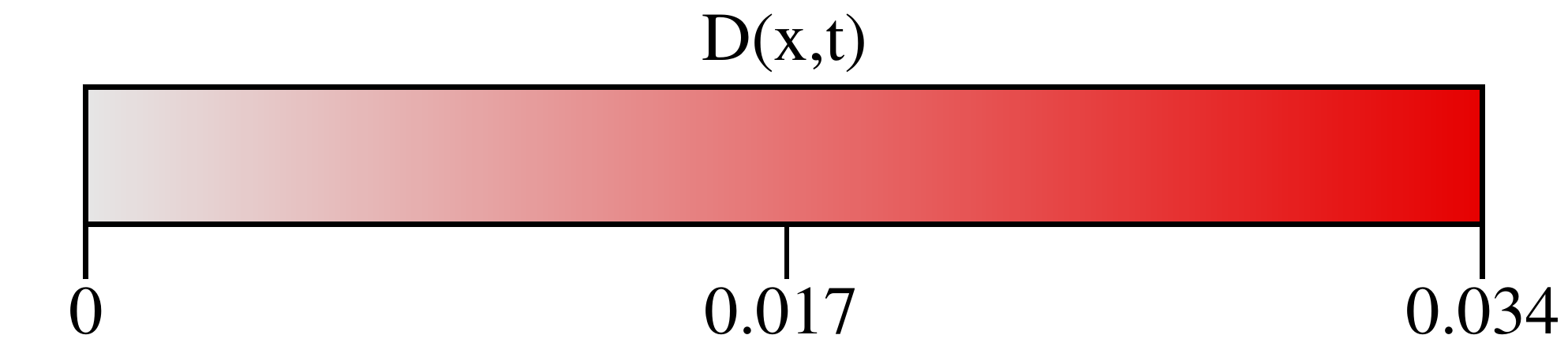}
}\\
\subfloat[$t=-2\,\hbar/U_0$]{\includegraphics[trim=0 100 0 0, clip, width=0.3\linewidth]{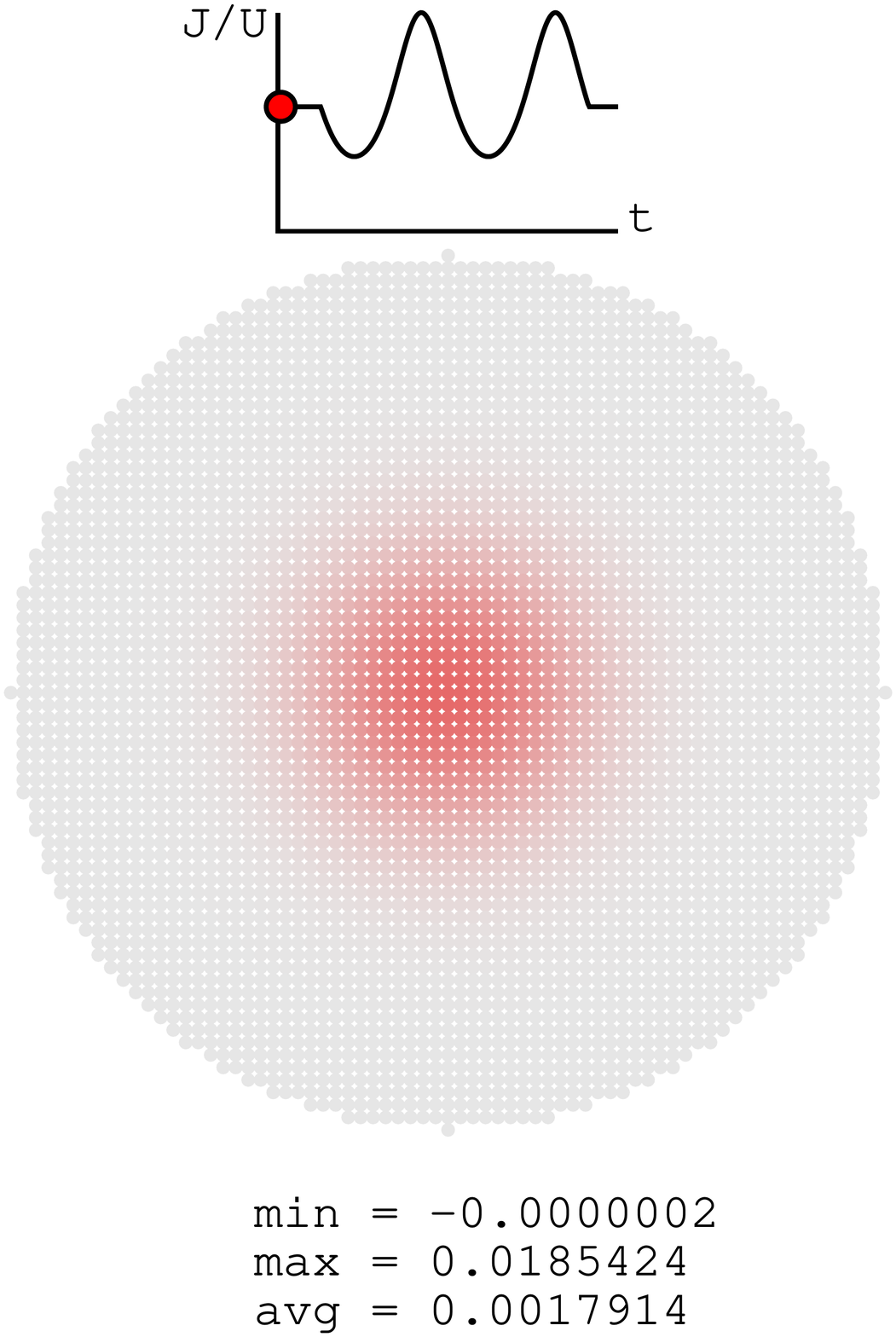}}
\subfloat[$t=1.23\,\hbar/U_0$]{\includegraphics[trim=0 100 0 0, clip, width=0.3\linewidth]{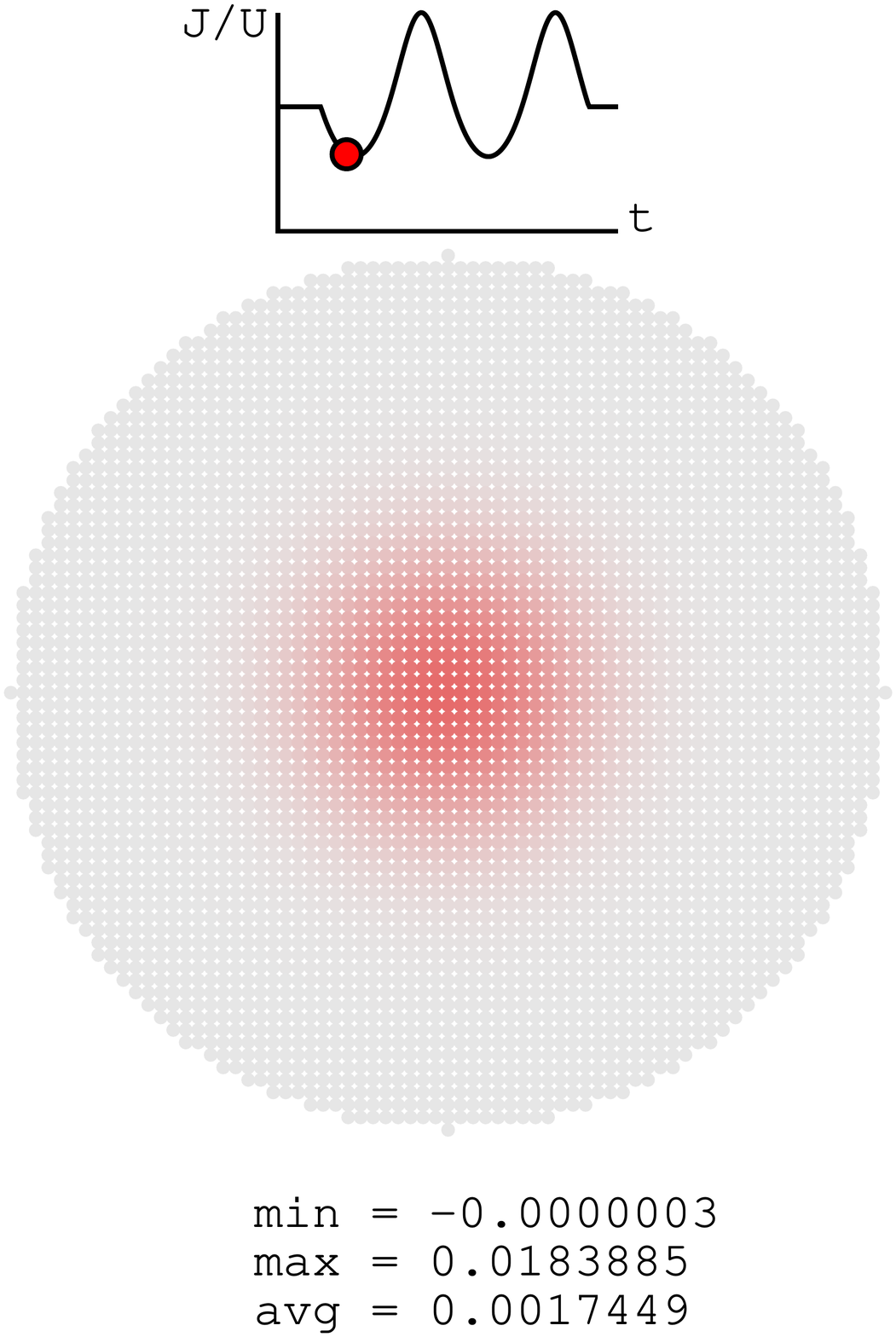}} 
\subfloat[$t=4.45\,\hbar/U_0$]{\includegraphics[trim=0 100 0 0, clip, width=0.3\linewidth]{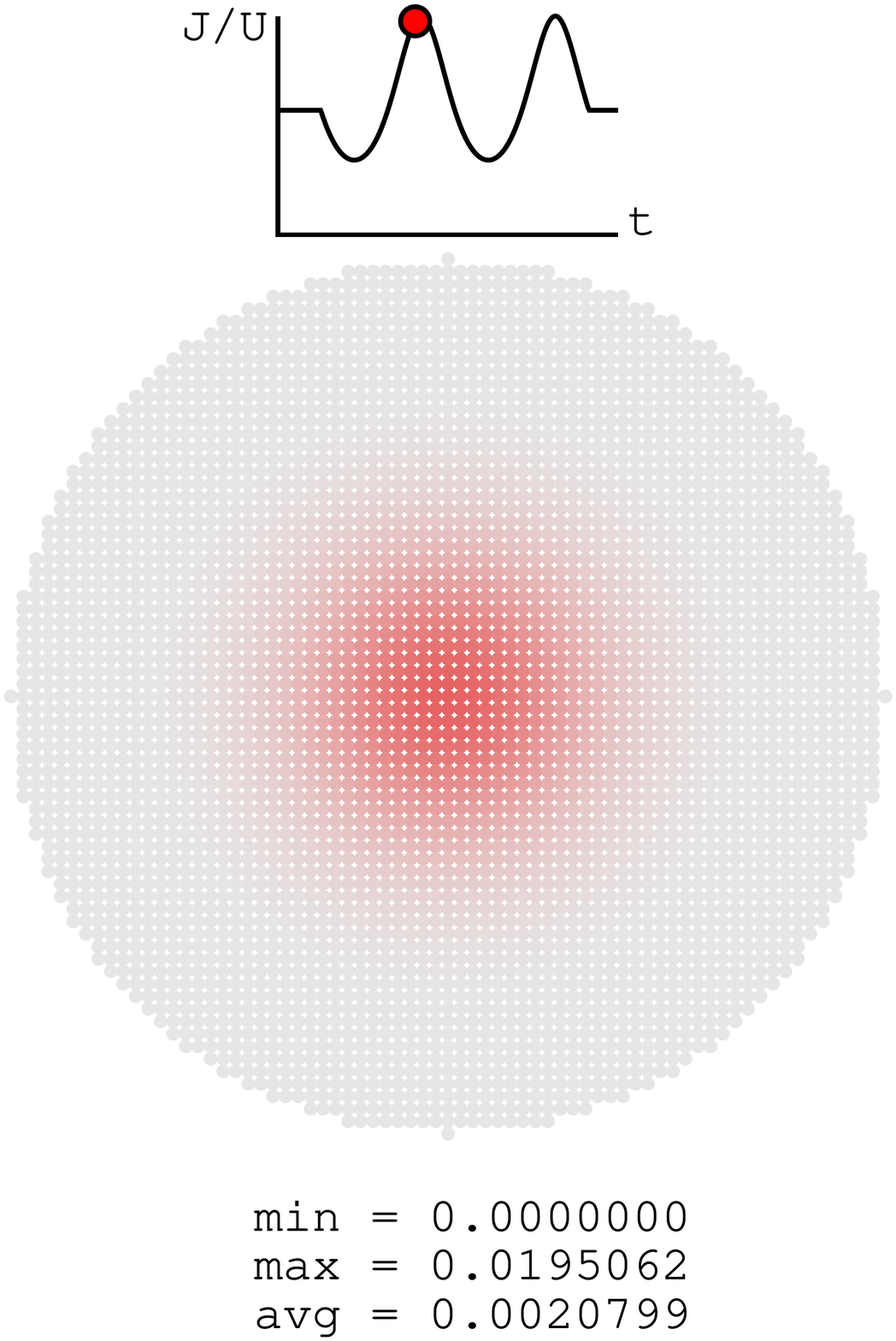}} \\
\subfloat[$t=7.68\,\hbar/U_0$]{\includegraphics[trim=0 100 0 0, clip, width=0.3\linewidth]{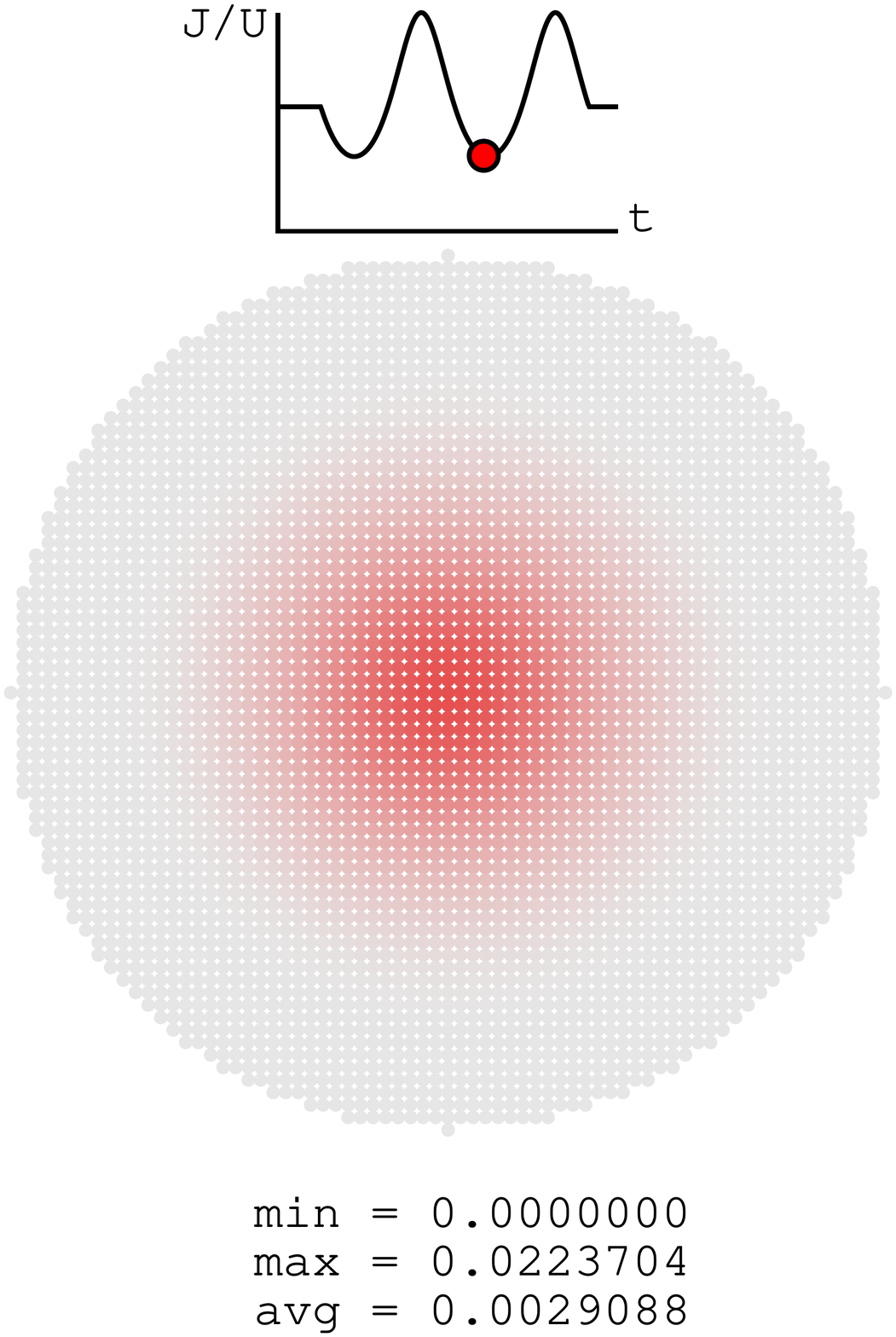}}
\subfloat[$t=10.90\,\hbar/U_0$]{\includegraphics[trim=0 100 0 0, clip, width=0.3\linewidth]{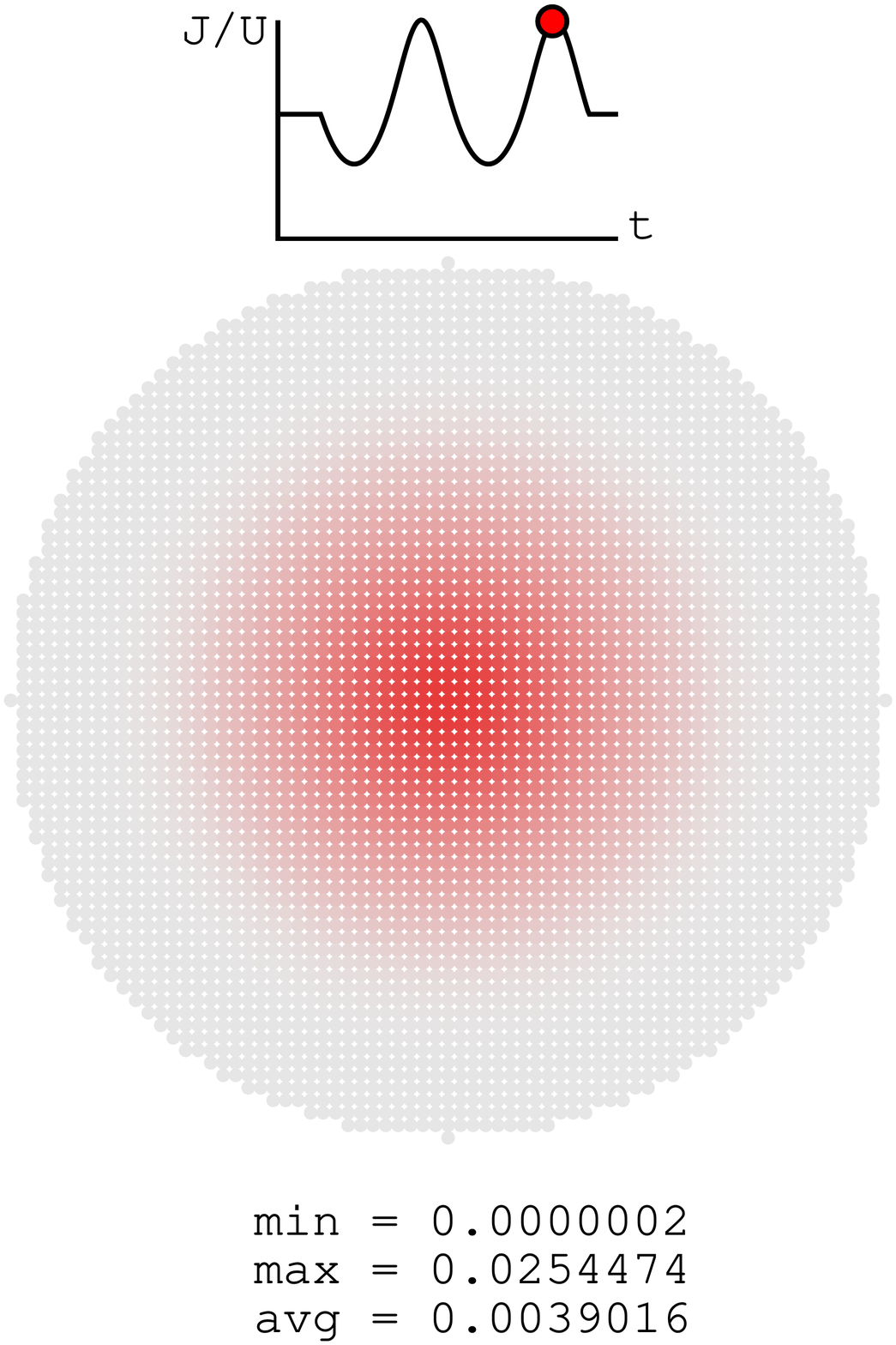}}
\subfloat[$t=14\,\hbar/U_0$]{\includegraphics[trim=0 100 0 0, clip, width=0.3\linewidth]{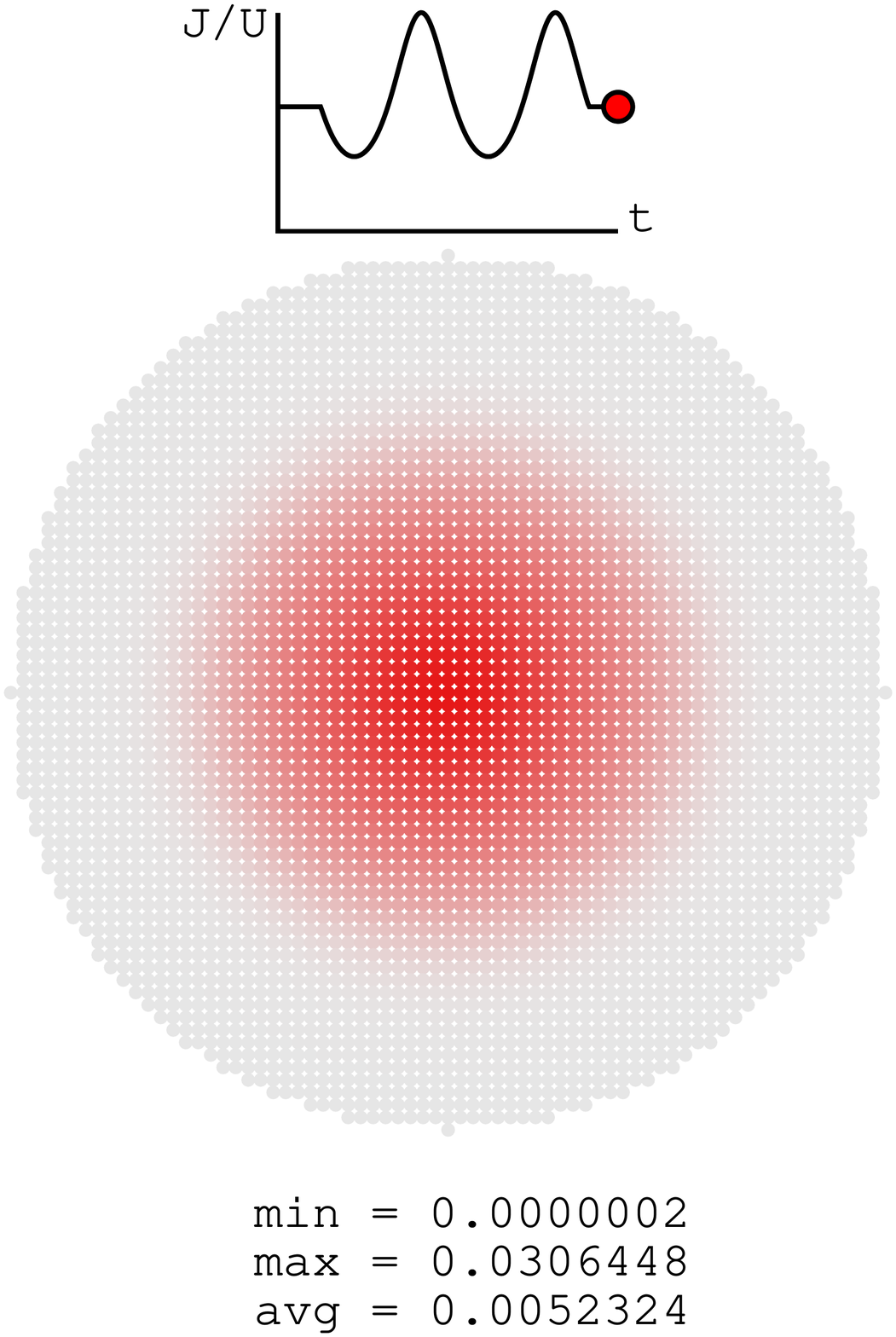}} \\
\subfloat[radial distributions]{\includegraphics[width=0.5\linewidth]{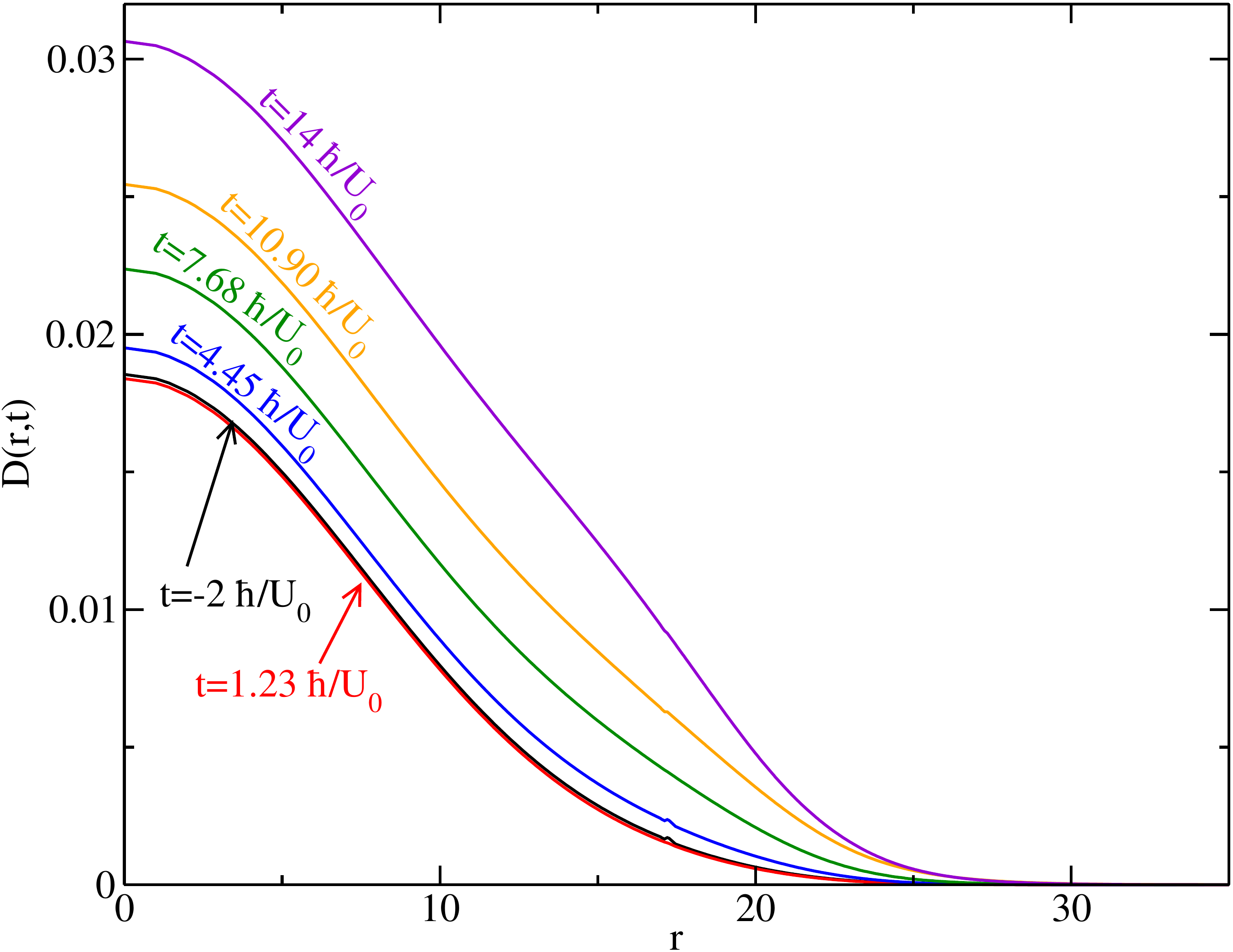}}
\caption{(color online) Direct numerical results for the double occupancy distribution $D(\vec x, t)$ as a function of time for a uniformly 
modulated lattice depth in a trap.
The insets of each panel show the hopping in units of the interaction $J(t)/U(t)$. 
The trap curvature is specified by $\rho_\text{trap}=4\,\text{sites}$.}
\label{fig:trapresults}
\end{figure*}

As a test system, we set the lattice depth to $V_0=10E_R$ and modulate it with an amplitude $\frac{\Delta V}{V_0}=20\%$ at 
the resonant frequency $\hbar \omega =U_0$. The interaction strength is chosen to be $U_0/6J_0=7.77$, and we assume an initial temperature
$k_BT=0.15U_0$. We choose to study two cycles of the modulation, that is $t_\text{mod}=2h/U_0$. For the simulations, we use up to 1024 time
slices and a lattice with up to 1024 symmetry-irreducible lattice sites, that is
up to 7844 actual lattice sites. The computational effort for a system with 512 symmetry-irreducible sites and a maximum of 1024 time slices 
is approximately $5\times 10^5$ CPU-hours on a Cray XE6. For instance, this involves 32768 CPUs for approximately 12 hours by the main 
simulation and some further CPU time for the cheaper simulations at larger $\Delta t$ which are required for the extrapolation $\Delta t \to 0$.

Figure \ref{fig:trapresults} shows simulation results for distribution of the double occupancy in a trapped system with 
$\rho_\text{trap}=4\,\text{sites}$ and the global chemical potential $\mu=0$. Each subfigure displays the distribution at a different point in time. 
Due to the lattice depth modulation,
the hopping in units of the interaction $J(t)/U(t)$ drives the system. The increases in the double occupancy occur as $J(t)/U(t)$ is decreasing.

\begin{figure}
\includegraphics[width=\linewidth]{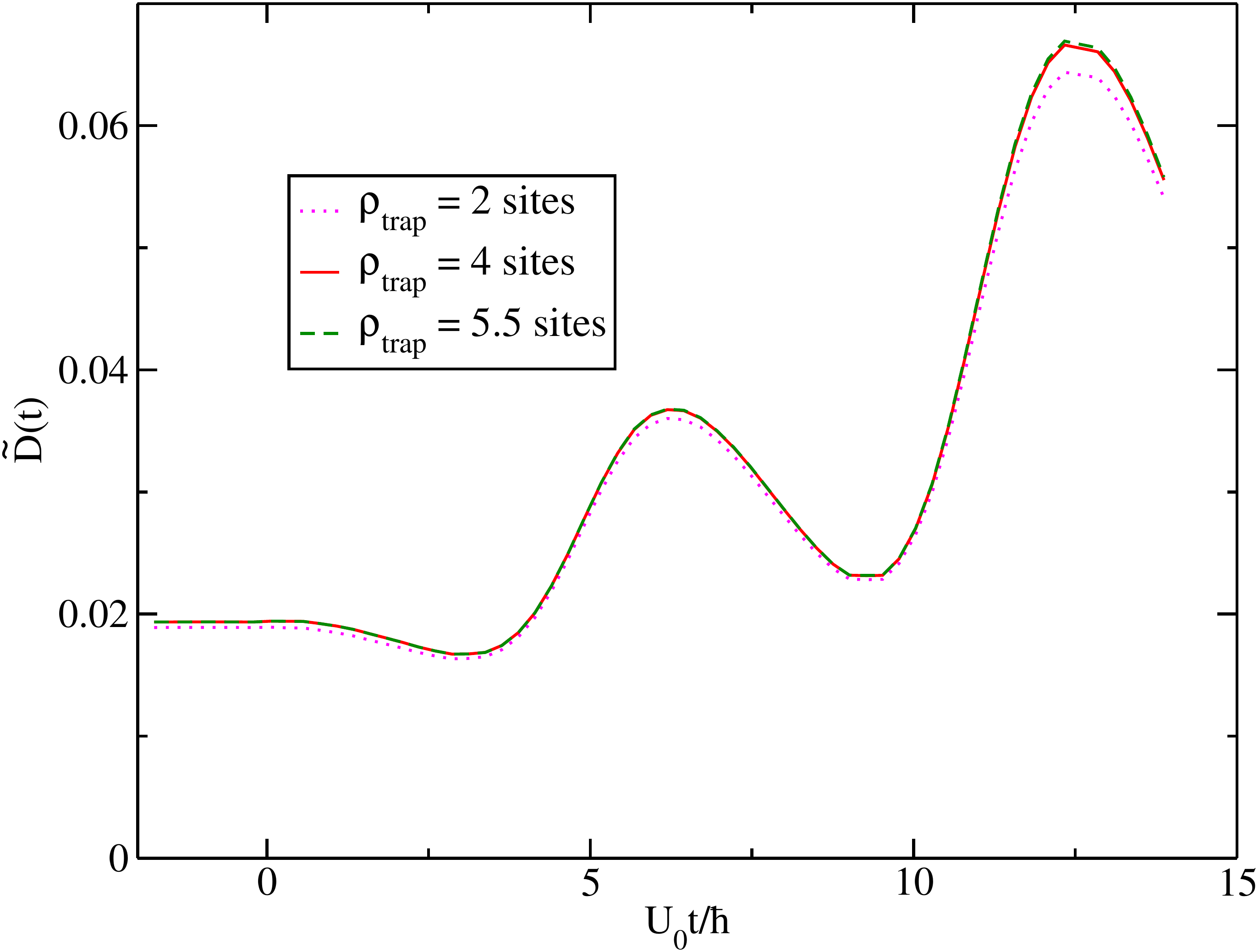}
\caption{(color online) Fraction of atoms on doubly occupied sites as a function of time for different trap curvatures.}
\label{fig:integrateddocc}
\end{figure}

To provide a better picture of the time dependence, Fig.~\ref{fig:integrateddocc} shows the fraction of atoms on doubly occupied sites,
\begin{equation}
\tilde D(t) = 
\frac{
2\sum_i \langle n_{i\uparrow} n_{i\downarrow} \rangle(t)
}
{
N
}
,
\end{equation}
where $N=\sum_i \langle n_i \rangle (t)=\text{const.}$
as a function of time for several values of the trap curvature. 
In the 
cases $\rho_\text{trap}=4\,\text{sites}$ and $\rho_\text{trap}=5.5\,\text{sites}$, the results lie on top of each other, whereas for 
$\rho_\text{trap}=2\,\text{sites}$ a slight deviation occurs.
This agrees with the results found in our previous publication Ref.~\cite{dirks2013} for homogeneous systems.

\subsubsection{Comparison to LDA}

In order to perform the comparison to the LDA, we solve numerous mutually independent homogeneous versions of the problem at several chemical
potentials $\mu=-V_\text{trap}(\vec r)$ and  compare to local observables obtained from the full trap simulation at a position $\vec r$.

We consider a set of test systems with three different trap curvatures, that is different values of the characteristic length
$\rho_\text{trap}$ of the trap potential, namely $\rho_\text{trap}=2\,$sites, $\rho_\text{trap}=4\,$sites, and $\rho_\text{trap}=5.5\,$sites.
In the simulations, the real part of the Kadanoff-Baym-Keldysh contour extends over an interval $[t_0, t_\text{max}]=[-2\hbar/U_0, 14 \hbar/U_0]$, whereas
the modulation acts over the interval $[0,t_\text{mod}]$. 

\begin{figure}
\includegraphics[width=\linewidth]{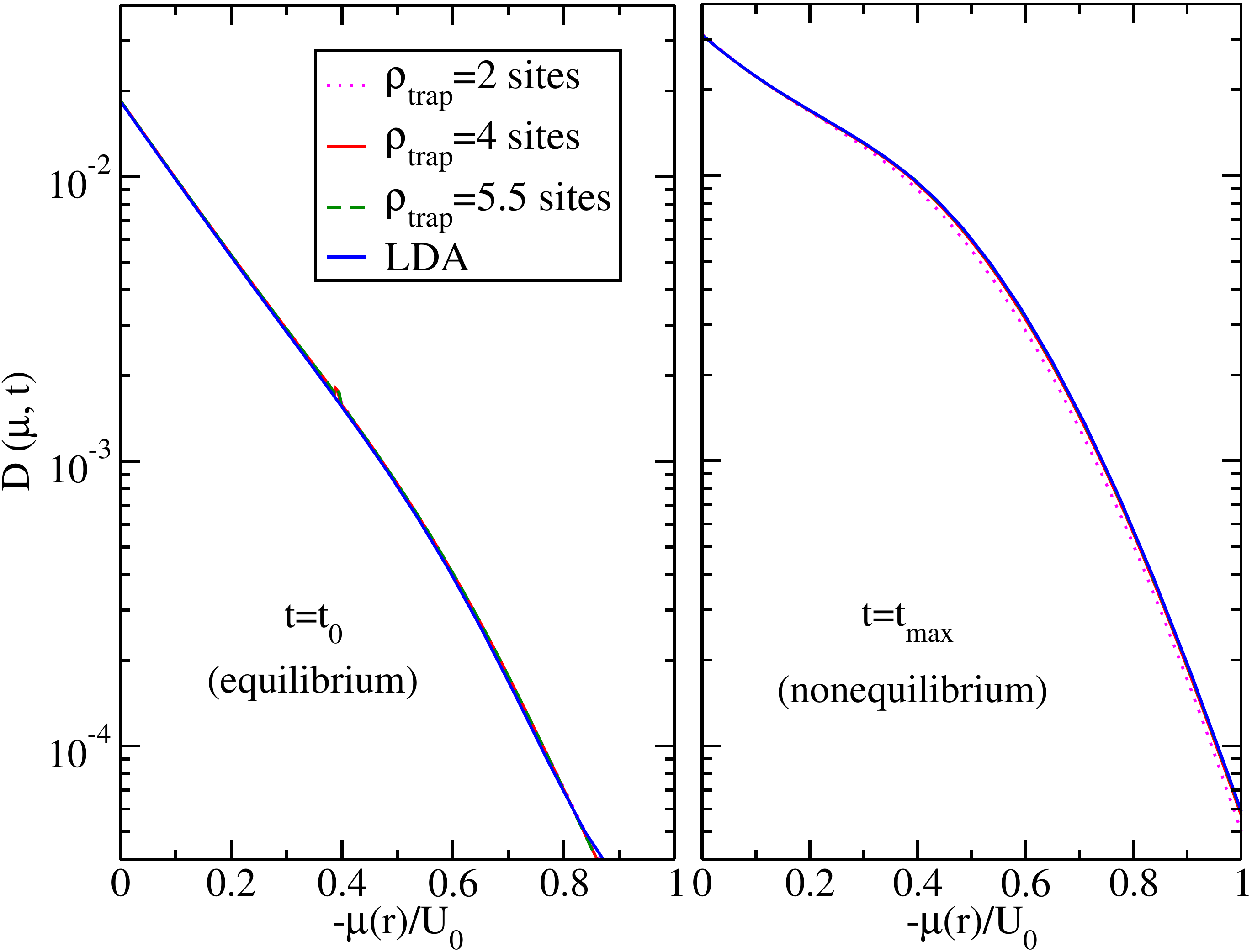}
\caption{(color online) Comparison of the double occupancies computed for the full trap simulation of traps with different curvatures as compared to the LDA result.}
\label{fig:LDAcompadocc}
\end{figure}

Figure \ref{fig:LDAcompadocc} shows a comparison of the double occupancy $D(\mu, t)$ as a function of the initial chemical potential $\mu$ 
at times before ($t=t_0$) and after ($t=t_\text{max}$) it has been driven out of equilibrium by the lattice depth modulation. 
As we see for both, the equilibrium ($t=t_0$) and nonequilibrium ($t=t_\text{max}$) situations, the numerical results for the 
inhomogeneous system agree well with the LDA, even for the rather steep trap potential with $\rho_\text{trap}=2\,$sites. 
The slight deviation of the solution for $2\,$sites from the other nonequilibrium curves may still be due to numerical imperfections.
The agreement with the LDA indicates that the creation of doubly occupied sites in a Mott insulator subject to a modulated lattice depth
is caused by strongly local excitation processes.

\section{Conclusion}
\label{sec:conclusion}
We presented a computational approach to an inhomogeneous Mott-insulating system of ultracold atoms. A major challenge is to
compute a large matrix inverse in the Dyson equation. We show that a GMRES-based inversion approach exploiting
the small numerical value of the hopping as compared to the many-body interaction yields a feasible implementation on 
supercomputers. A comparison to the LDA shows that both methods are well-suited for the problem of lattice-depth modulation 
spectroscopy. This hints towards mainly local processes being involved in the coherent excitations between lower and upper Hubbard bands
in this particular setting, as might have been anticipated.
In the future, we will apply the inhomogeneous method to problems with mass transport where the LDA is expected to fail.

At present, the computational complexity of the algorithm is proportional to $N_t^4$. It may be worthwhile to investigate the possibility
to extend the time range by truncating certain parts of the self-energy at a given threshold for $t-t'$. This measure could increase the
applicability of the algorithm greatly but requires further efforts. 

\section{Acknowledgments}
This work was supported by a
MURI grant from the Air Force Office of Scientific Research
numbered FA9559-09-1-0617. Supercomputing resources
came from a challenge grant of the DoD at the Engineering Research
and Development Center and the Air Force Research and
Development Center. The collaboration was supported
by the Indo-US Science and Technology Forum under
the joint center numbered JC-18-2009 (Ultracold atoms).
JKF also acknowledges the McDevitt bequest at Georgetown.
HRK acknowledges support of the Department of 
Science and Technology in India.

\appendix
\section{Utilization of Symmetries}
\label{app:symmetries}
\begin{figure}
\centering
\includegraphics[width=0.7\linewidth]{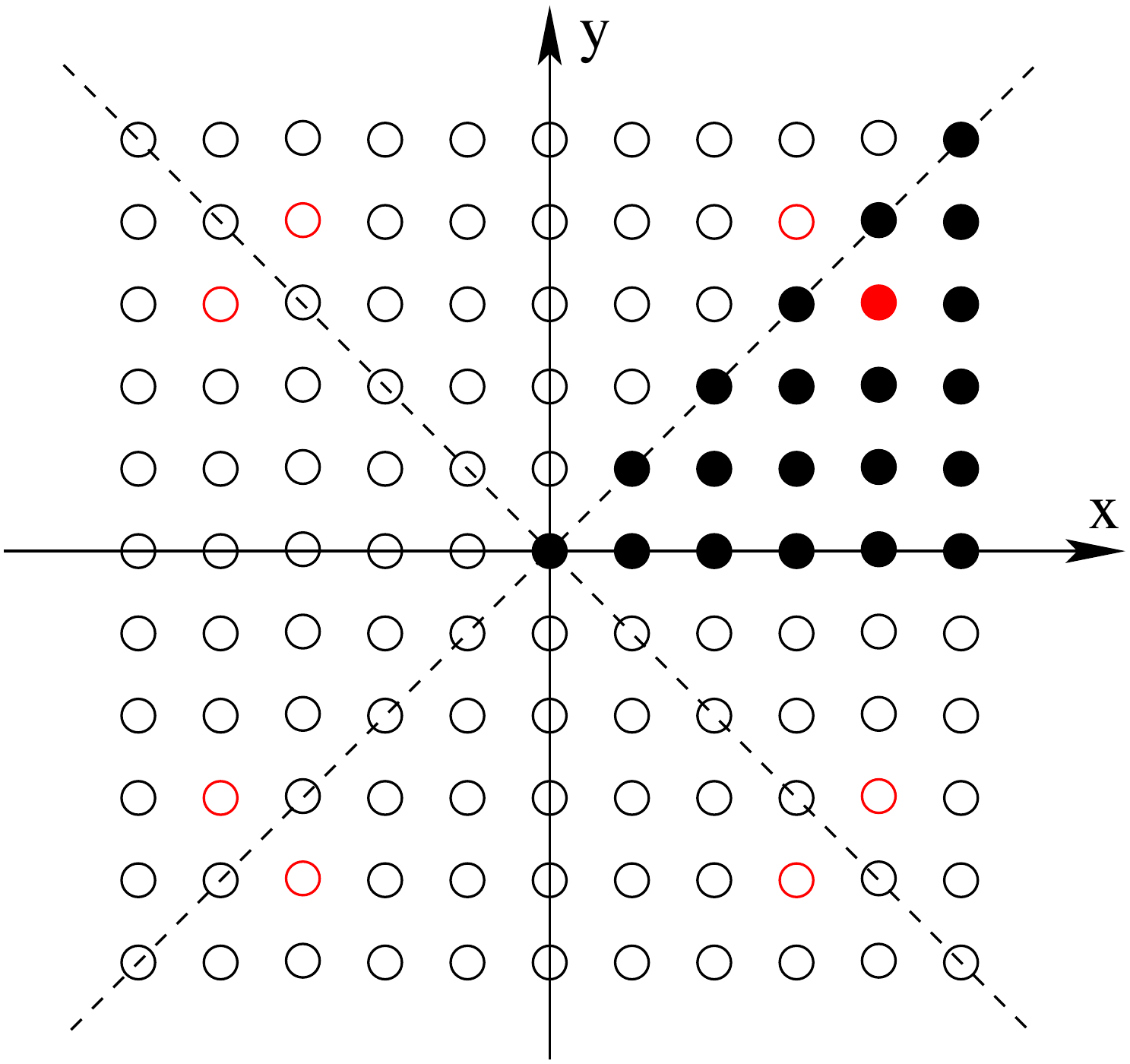}
\caption{(color online) $C_{4v}$ point-group symmetry imposed by the circular trap potential on the 2D lattice. 
The circles denote lattice sites. Solid circles are representatives of an
equivalence class of lattice sites with respect to the symmetry. The symmetry partners of the red/gray solid circle
are displayed in red/gray.}
\label{fig:symmetry}
\end{figure}
We illustrate the exploitation of symmetries for the example of spherical symmetry on a two-dimensional square lattice with an $s$-orbital 
basis. In this case, a $C_{4v}$ point group symmetry \cite{cotton} is imposed on the lattice.
Fig.~\ref{fig:symmetry} displays the lattice sites of a lattice. For the $C_{4v}$ symmetry, 
all sites can be represented by the sites in the irreducible wedge $x\geq 0 \wedge y \leq x$. These representatives of the equivalence classes 
are denoted by solid black circles.
If the Green's function and self-energy transform as the identity representation of the point group, the computations only need to be performed for those representatives.
The representative of a given site can be retrieved by reflections with respect to the coordinate axes and the diagonals which are shown as dashed
lines in Fig.~\ref{fig:symmetry}.

\section{Formula for the double occupancy}
\label{app:doccformula}
We provide a brief derivation of Eq.~\eqref{eq:docc}.
Let us start by assuming an equidistant discretization $\{t_0, \dots, t_N\}$ ($\Delta t = t_{i+1}-t_i$) of the forward part of 
the Kadanoff-Baym-Keldysh contour. We obtain
\begin{equation}
\begin{split}
\left.\frac{\partial G_{l\sigma} (t,t')}{\partial t} \right|_{t'=t^+} = &
\frac{1}{\Delta t}  (G_{l\sigma}(t_{i}, t_i) - G_{l\sigma} (t_{i-1}, t_i)) \\
& + \mathcal{O}(\Delta t) \\
=& \frac{-\imag}{\Delta t}\Big\langle \euler{\imag H(t_0)\Delta t} \cdots \euler{\imag H(t_{i-1})\Delta t}\,\,  \\
& \qquad\qquad\qquad \times\,\, \hat A_{l\sigma}\,\, \\
& \quad\times \,\euler{-\imag H(t_{i-1})\Delta t}\cdots  \euler{-\imag H(t_{0})\Delta t}\Big\rangle \\
& + \mathcal{O}(\Delta t), \\
\end{split}
\end{equation}
with
\begin{equation}
\hat A_{l\sigma} = -\euler{\imag H(t_i) \Delta t}c^\dagger_{l\sigma}\left[ c_{l\sigma}, \euler{-\imag H(t_i)\Delta t}  \right].
\end{equation}
The operator $\hat A_{l\sigma}$ simplifies as follows:
\begin{equation}
\begin{split}
\hat A_{l\sigma} =& -\imag \Delta t \cdot c_{l\sigma}^\dagger[H(t_i), c_{l\sigma}]  + \mathcal{O}(\Delta t^2) \\
 = & -\imag \Delta t \cdot \bigg( \sum_j J_{lj} c^\dagger_{l\sigma}c_{j\sigma} - \epsilon_{l\sigma}(t) n_{l\sigma} - \\
   & \qquad\quad - U_l(t) n_{l\uparrow} n_{l\downarrow}\bigg) + \mathcal{O}(\Delta t^2).
\end{split}
\end{equation}
Taking the limit $\Delta t\to 0$ results in Eq.~\eqref{eq:docc}. Numerically, this limit must be performed via linear and/or quadratic extrapolation
of multiple simulations for different $\Delta t$ values.

\section{GMRES}
\label{app:gmres}
The Generalized Minimal Residue Method was introduced by Saad and Schultz \cite{GMRES} to solve a linear equation
\begin{equation}
\hat A x = b.
\label{eq:GMRES1}
\end{equation}
A good introduction to the method can be found in Ref.~\cite{gmreslecturenotes}, and a useful C++ implementation is provided by the
NIST IML++ template library \cite{imllibrary}.
In order to solve the equation \eqref{eq:GMRES1}, GMRES operates in $d$-dimensional Krylov subspaces
\begin{equation}
\mathcal{K}_d(\hat A, r^{(0)}) = \text{span}\,(
r^{(0)}, \hat A r^{(0)}, \dots, \hat A^{d-1}  r^{(0)}
    )
\end{equation}
which are successively built up, starting with $d=1$. $r^{(0)} = b- Au^{(0)}$ is the residue of the initial guess $u^{(0)}$ for the solution.
The GMRES method forms a refined approximate solution $u^{(d)}\in u^{(0)} + \mathcal{K}_d(\hat A, r^{(0)})$ defined by the minimization 
requirement
\begin{equation}
u^{(d)} \in u^{(0)} + \mathcal{K}_d(\hat A, r^{(0)}), \quad\| b - \hat A u^{(d)} \|_2 \mathop{=}^{!} \text{min},
\end{equation}
where the expression "$x\mathop{=}^{!} \text{min}$" demands $x$ to be minimal.
The Krylov space dimension $d$ is increased until the convergence criterion \eqref{eq:GMRESconvergence} is reached or $d$ 
approaches the threshold $d_\text{max}$ at which the minimization is considered too expensive. If $d=d_\text{max}$, 
GMRES is restarted with a Krylov space size $d=1$, where the initial guess 
$u^{(0)}$ is taken to be $u^{(d_\text{max})}$ from the previous GMRES iteration before restarting.
With a preconditioner $\mathcal{P}$, one applies GMRES to the system
\begin{equation}
\mathcal{P}\hat A x = \mathcal{P} b,
\label{eq:GMRESPreconditioned}
\end{equation}
rather than Eq.~\eqref{eq:GMRES1}.

If $\mathcal{P} \hat A \approx \hat I$ this system is better conditioned than Eq.~\eqref{eq:GMRES1}.
In the case that $\hat A$ is a sparse matrix with large entries on the diagonal and some randomly occuring off-diagonal entries, 
the diagonal matrix $\hat B$ defined by the diagonal entries of $\hat A$ yields a good preconditioner $\hat B^{-1}$, because the condition number 
\begin{equation}
\kappa = \frac{\sigma_\text{max}(\hat A)}{\sigma_\text{min}(\hat A)}
\end{equation}
is not as close to $1$ as the regularized
\begin{equation}
\tilde \kappa = \frac{\sigma_\text{max}(\hat B^{-1}\hat A)}{\sigma_\text{min}(\hat B^{-1}\hat A)}.
\end{equation}
Here, $\sigma_\text{max}$ and $\sigma_\text{min}$ represent the respective maximal and minimal singular values.
Also, the application of the diagonal matrix $\hat B$ to a vector is a cheap operation.
The same argument can be made in the case that $\hat A$ has large entries on the block diagonal $\hat B$, as it is the 
case in the nonequilibrium inhomogeneous strong-coupling expansion, where $\hat A = \hat B-\hat J$.
The block diagonal $\hat B$ is defined in Eq.~\eqref{eq:definitionB}, and the hopping matrix $\hat J$ is a sparse matrix with small numerical values
(see also Eq.~\eqref{eq:Ginvblockstruct}).
Due to the latter, in fact $\hat B^{-1} \hat A \approx \hat I$, so the equation system is well-conditioned.

The actual GMRES algorithm with preconditioner in pseudo code reads \cite{gmreslecturenotes, imllibrary}
\begin{enumerate}
\item For the initial guess $u^{(0)}=\hat B^{-1}b$ compute the preconditioned residue $z^{(0)} = \hat B^{-1} (b-\hat A u^{(0)})$, as well as $q^{(1)} = z^{(0)} / \|z^{(0)}\|_2$.
Initialize the Hessenberg matrix
\begin{equation}
H = (h_{ij})_{\substack{1\leq i \leq d_\text{max}+1\\1\leq j\leq d_\text{max}}} = 0.
\end{equation}
\item
For $d = 1, \dots, d_\text{max}$ do
\begin{itemize}
\item[]$ w = \hat B^{-1}\hat A q^{(d)}$ 
\item[]For $i = 1, \dots, d$ do 
\item[]
\begin{itemize}
\item[]$h_{id} = \left\langle q^{(i)}| w\right\rangle$
\item[]$w \to w - h_{id} q^{(i)}$
\end{itemize}
\item[]$h_{d+1,d} = \|w\|_2$
\item[]If $h_{d+1,d} = 0$ then proceed with step 3 to compute the result.
\item[]Otherwise use step 3 to check for convergence, Eq.~\eqref{eq:GMRESconvergence}. Continue if not converged.
\item[]$q^{(d+1)}=w/h_{d+1,d}$
\end{itemize}
\item
Solve the $d$-dimensional linear minimization problem
\begin{equation}
\|\|z^{(0)}\|_2e_1 - H^{(d)} y\|_2 \to \text{min},
\end{equation}
where $H^{(d)}$ is the upper left $d$-dimensional square of $H$,
to obtain the result $y^{(d)}$.\\
Then set $u^{(d)}=u^{(0)} + Q^{(d)} y^{(d)}$, with $Q^{(d)} = (q^{(1)} \cdots q^{(d)})$.
\\
\end{enumerate}

Practically, the GMRES with preconditioner scans for solutions in the modified affine Krylov spaces
\begin{equation}
\begin{split}
&u^{(0)} + \mathcal{K}_d(\hat B^{-1}\hat A, z^{(0)})\\
&=\hat B^{-1} b +
\text{span}\,(
v_1, \dots, v_d
    ).
\end{split}
\end{equation}
Here, the basis vectors are
\begin{equation}
v_n = (\hat B^{-1}\hat A)^{n-1}  \hat B^{-1}(b-\hat A \hat B^{-1}b)).
\end{equation}
In the case of the inhomogeneous nonequilibrium strong-coupling expansion, $b$ is a unit vector localized at a given lattice site.
The $n$-th basis vector $v_n$ of the Krylov space $\mathcal{\tilde K}_d(\hat B^{-1}\hat A, z^{(0)})$ corresponds to $n$ hopping 
processes, because $\hat A = \hat B - \hat J$, and in $v_n$, $\hat A$ is applied $n$ times to $b$.
That is, the GMRES method includes exactly $d$ iterated hopping processes in iteration $d$ until 
convergence is reached. If GMRES is restarted, the $d$'th iteration includes $md_\text{max}+d$ iterated hopping
processes, where $m$ is the number of restarts.
This is crucial for the computational optimizations used in the implementation.
The procedure can be seen as a numerically controlled analogue to a partial summation of the Dyson series
\begin{equation}
\frac{1}{\hat B-\hat J}b = \sum_{\nu=0}^\infty(\hat B^{-1} \hat J)^\nu \hat B^{-1}b.
\end{equation}

\end{document}